\documentclass[twocolumn,amsmath,amssymb,prb,a4paper,showpacs]{revtex4}
\usepackage{graphicx}
\usepackage{dcolumn}
\usepackage{bm}

\newcommand{\ket}[1]{| #1 \rangle}
\newcommand{\bra}[1]{\langle #1 |}
\newcommand{\braket}[2]{\langle #1 | #2  \rangle}

\newcommand{\U}{\mathbf{U}}
\newcommand{\V}{\mathbf{V}}

\newcommand{\NL}{N_{\mathrm{L}}}
\newcommand{\NR}{N_{\mathrm{R}}}
\newcommand{\NT}{N_{\mathrm{T}}}

\newcommand{\WL}{\mathbf{W}_{\mathrm{L}}}
\newcommand{\WR}{\mathbf{W}_{\mathrm{R}}}
\newcommand{\SL}{\bm{\Sigma}_{\mathrm{L}}}
\newcommand{\SR}{\bm{\Sigma}_{\mathrm{R}}}

\newcommand{\LL}{\bm{\Lambda}_{\mathrm{L}}}
\newcommand{\LR}{\bm{\Lambda}_{\mathrm{R}}}

\newcommand{\W}{\mathbf{W}}
\renewcommand{\H}{\mathbf{H}}
\renewcommand{\S}{\bm{\Sigma}}

\newcommand{\DTM}{\Delta T_{max}}
\newcommand{\DTA}{\Delta T_{ave}}

\begin{document}
\title{{Determination of complex absorbing potentials from
the electron self-energy}}
\author{{Thomas M. Henderson, Giorgos Fagas\footnote{Corresponding author.
E-mail: gfagas@tyndall.ie}, Eoin Hyde, and James C.  Greer}}
\affiliation{
Tyndall National Institute, 	
Lee Maltings,
Prospect Row, 			
Cork, Ireland}
\date{\today}

\begin{abstract}
The electronic conductance of a molecule making contact to electrodes is 
determined by the coupling of discrete molecular states to the continuum 
electrode density of states.  Interactions between bound states and continua 
can be modeled exactly by using the (energy-dependent) self-energy, or 
approximately by using a complex potential.  We discuss the relation between 
the two approaches and give a prescription for using the self-energy to 
construct an energy-independent, non-local, complex potential.  We apply our 
scheme to studying single-electron transmission in an atomic chain, obtaining 
excellent agreement with the exact result. Our approach allows us to treat
electron-reservoir couplings independent of single electron energies, allowing
for the definition of a one-body operator suitable for inclusion into correlated
electron transport calculations.
\end{abstract}

\maketitle

\section{Introduction}
\label{sec:Intro}
Initiated by experimental advances, interest has been growing in the
first-principles description of quantum transport through nanojunctions formed
by a single-molecule bridge between electrodes acting as electron reservoirs 
(see Ref.~\onlinecite{Book} for a recent overview).  Molecular electronic 
structure, including its response to external fields, are well-described by 
{\it ab initio} methods if many-electron methods are used. 
Recent studies also point to the need for 
a detailed quantum chemical approach to predict current-voltage characteristics
for electron transport across single 
molecules~\cite{PRL03DG,PRB06FDG,PRB06KBE,JAP06ASS,TR06}.
Unfortunately, the interaction of the molecular energy states with the bulk 
electrode density of states makes explicit treatment of the many-electron problem
intractable.

Similar coupling of bound and continuum states occurs in many diverse
processes in chemistry and physics but a completely first principles description 
is generally difficult.  For example, with conventional, 
basis-set-dependent methods, it is not feasible to efficiently describe both 
bound and continuum states simultaneously.  However, it has been known for 
some time that a reduced description for a finite number of degrees of freedom 
can be formally achieved by adding to the uncoupled Hamiltonian of the 
selected subsystem a non-Hermitian effective 
interaction~\cite{SantraCederbaumCAP,MugaCAP,PR98Moiseyev,PR91Domcke}.
As a result of this extra term, the line spectrum of the uncoupled Hamiltonian 
evolves so that the sharp energy levels become resonances with broadenings and 
shifts that depend on the form and strength of the coupling to the continuum.

An exact non-Hermitian interaction can be obtained through the
Feshbach-Fano~\cite{Feshbach1,Feshbach2,PR61Fano} projection operator 
technique and is essentially the self-energy of Green's function
methods~\cite{PR91Domcke,PRL59BS,AP60FN,PRL00Ceder,JCP02SC,JCP03FSSC}.
It is not entirely clear how best to proceed in the general case but there are 
now standard numerical methods to extract a first-principles self-energy for 
bulk electrodes in the single-particle approximation~\cite{Book}.  For certain 
model systems, analytical expressions are
available~\cite{GiorgosHuckelCAP2,CP06MR}. If we could use
this coupling to the reservoirs, we could describe quantum 
transport through a nanojunction using sophisticated many-body methods on a region
(e.g., an ``extended'' molecule including the molecule plus some part of the leads)
while retaining a single particle description of the bulk electrodes.
This would open up the possibility to
improve on the commonly employed but controversial single-particle description 
of single-molecule devices~\cite{PRB06FDG,PRB06KBE,JAP06ASS,TR06}.
However, application of one-electron self-energies in many-electron calculations is not
straightforward, as the single particle levels coupled to reservoirs through the
self-energy terms have no immediate corresponding quantities within correlated
electronic theories. Hence we seek a method that is able to transmit the
information contained within the self-energies, e.g., the coupling to
electron-reservoirs, directly to a many-electron description.

The formally exact complex-scaling technique provides just such an
alternative~\cite{PR98Moiseyev}.  By applying a complex-coordinate 
transformation to the Hamiltonian via smooth-exterior-scaling the 
Schr{\"o}dinger equation is transformed; the eigenvalues remain the same but 
the wave functions obey different boundary conditions.  This yields exact 
resonant positions and widths. However, for the specific
applications we have in mind it is operationally more convenient to use another
general means that is commonly employed and only requires the addition of an
energy-independent, phenomenological complex potential to the
Hamiltonian~\cite{SantraCederbaumCAP,MugaCAP}.

These potentials are typically local in space and purely imaginary, with negative
imaginary part; they vanish inside the subsystem region (well-described by the
usual basis set), and grow rapidly away from that 
region.  The negative imaginary walls cause asymptotic damping of resonant 
eigenfunctions, preventing them from extending to infinity.  The wave function 
becomes square-integrable, obviating the need to describe free-particle states 
or the associated continuum.  Hence, standard bound-state methods can be 
applied to open systems.  Because these potentials effectively absorb 
particles that would otherwise escape to infinity, they are known as complex 
absorbing potentials (CAPs).

Calculations may depend rather sensitively on the detailed form of the
CAP~\cite{JCP92SM,JCP95Neu,CPL04MN,CPL94MBM,RissMeyerCAP2}, and one
must therefore be careful in constructing and parameterizing the complex 
potential.  Semi-classical arguments have been used to suggest a 
form~\cite{JCP92VK,JCP02Manol,JCP03PC}, and constraints on the form are 
known~\cite{RissMeyerCAP1}. Parameters are sometimes fit to 
experiment; more frequently they are numerically optimized
so that the stationarity condition for the complex-value of the resonant energy
is satisfied.~\cite{RissMeyerCAP1}.  Deviations occur
because in practice complex potentials not only absorb particles but also
cause artificial reflections. In fact though, Riss and
Meyer~\cite{RissMeyerCAP3} and Moiseyev~\cite{JPB98Moiseyev} pointed out that
there is a relation between complex-scaling and CAPs in the limit of zero
reflections which may be used to introduce exact reflection-free CAPs;
the parameters in their functional form are determined by further stability
conditions~\cite{JPB98Moiseyev}.

While the utility of the complex potential approach has been demonstrated many
times~\cite{SantraCederbaumCAP,MugaCAP}, we require a technique which allows us to
avoid a search over parameter space if we are to study molecules interacting with
large electron reservoirs.
Further, we would like to make use of the well-understood
techniques for calculating the first-principles self-energy for bulk electrodes.
This suggests relating the self-energy with an energy-independent complex
absorbing potential.  This CAP could then be used in a many-body calculation
as an additional one-body potential, similar to other applications of complex
absorbing potentials~\cite{CPL99SCM,CP06SSM}. This would allow us to apply
many-body methods on a region while using the self-energy transformed CAP to
couple to the reservoirs.

In this paper, our purpose is to present a method to determine a CAP starting
from the self-energy. As as first application of our approach we calculate
the transmission spectrum of independent electrons propagating through a model
electrode-molecule-electrode junction that has been previously reported
from calculations using both exact
methods~\cite{GiorgosHuckelCAP1,GiorgosHuckelCAP2} and a numerically optimized 
CAP~\cite{KopfSaalfrank}.  Remarkable agreement with the exact result is 
found, thus substantiating a link between the exact self-energy and an 
energy-independent CAP.

The structure of the paper is as follows.  In Sec.~\ref{sec:Theory}, we 
introduce background theoretical material for the coupling of bound states to 
a continuum and derive our CAP from the self-energy.  
Section~\ref{sec:Application} summarizes the application of our method to 
single-electron transmission through a simple model for an 
electrode-molecule-electrode junction described in 
Sec.~\ref{sec:ModelSystem}.  The results of our calculations are presented in 
Sec.~\ref{sec:Results}.  We conclude with a summarizing section.

\section{Theory}
\label{sec:Theory}
In order to exploit the apparent relation between the self-energy and the complex
absorbing potential, we assume that an energy-independent Hamiltonian with the ``correct''
broadened and shifted energy levels and states contains all the relevant physics from the 
interaction to the continuum.  We must simply define these energy levels and 
states and then build the operator.

\subsection{Coupled States and Broadened Energies from the Self-Energy}
Suppose we have a bare Hamiltonian $H_0$ which describes a closed, finite 
system.  The fact that the system is closed and finite means that $H_0$ will 
have bound states $\ket{\chi_i}$ with sharp energies $\epsilon_i$.  That is, 
we have
\begin{subequations}
\begin{eqnarray}
H_0 \ket{\chi_i} &=& \epsilon_i \ket{\chi_i},
\\
\bra{\chi_j} H_0 &=& \epsilon_j \bra{\chi_j},
\\
\braket{\chi_i}{\chi_j} &=& \delta_{ij}.
\end{eqnarray}
\label{BareEquations}
\end{subequations}
If we now couple our system to a continuum (or, equivalently, give it infinite 
extent), the energy levels $\epsilon_i$ will become 
$\omega_i = \epsilon _i + \delta_i - \mathrm{i} \lambda_i$, where $\delta_i$ 
is a shift in the position of the energy level and $\lambda_i$ is a width, and 
the states described by $\ket{\chi_i}$ will change.  Our goal is to obtain 
these new energy levels and states.

This can be achieved in a formally exact way using the self-energy $\Sigma$ 
from Green's function theory; the proper states and energy levels can be 
obtained by solving the Dyson equation.  That is, we solve~\cite{note1}
\begin{subequations}
\begin{eqnarray}
\left[H_0 + \Sigma(\omega_i) \right] \ket{\psi_i} &=& \omega_i \ket{\psi_i},
\\
\bra{\phi_i} \left[H_0 + \Sigma(\omega_i) \right] &=& \omega_i \bra{\phi_i};
\end{eqnarray}
\label{Dyson}
\end{subequations}
the real part of $\omega_i$ gives the position of the $i^{\textrm{th}}$ 
resonance including the shift from $\epsilon_i$, and the imaginary part gives 
the level broadening.  There will almost always be more solutions of this 
equation than there were eigenvalues of our original bare Hamiltonian.  The 
extra solutions correspond to states dominated by the continuum to which we 
have coupled, and in these states we are not particularly interested.  We need 
some process, then, to find the states and energy levels that best correspond 
to the states and energy levels of the bare Hamiltonian.

To do this, we solve the related problem
\begin{subequations}
\begin{eqnarray}
\left[H_0 + \lambda \Sigma(\omega_i^{\lambda}) \right] \ket{\psi_i^{\lambda}} &=& \omega_i^{\lambda} \ket{\psi_i^{\lambda}},
\\
\bra{\phi_i^{\lambda}} \left[H_0 + \lambda\Sigma(\omega_i^{\lambda}) \right] &=& \omega_i^{\lambda} \bra{\phi_i^{\lambda}}.
\end{eqnarray}
\end{subequations}
At $\lambda = 0$, we have the original states $\ket{\chi_i}$ and energy levels 
$\epsilon_i$, while at $\lambda = 1$, we have the target states and energy 
levels.  We simply let $\lambda$ go adiabatically from 0 to 1, thereby 
obtaining the desired states $\ket{\psi_i}$ and $\bra{\phi_i}$ and energy 
levels $\omega_i$.

\subsection{Definition of the CAP Operator}
In principle, we would like to build an energy-independent complex potential 
$W$ such that the Hamiltonian $H_0 + W$ has the $\omega_i$ as eigenvalues, and 
the $\bra{\phi_i}$ and $\ket{\psi_i}$ as left- and right-hand eigenvectors.  
Unfortunately, we cannot actually construct such a Hamiltonian.  This is 
because if $\bra{\phi_i}$ and $\ket{\psi_j}$ are eigenvectors of the same 
Hamiltonian, they must satisfy $\braket{\phi_i}{\psi_j} = \delta_{ij}$.  But 
because we obtain $\bra{\phi_i}$ and $\ket{\psi_j}$ at different energies
$\omega_i \neq \omega_j$, they do not obey this biorthogonality relationship.  
We must therefore consider constructions which yield eigenvalues that are 
approximately the $\omega_i$ and eigenvectors that are approximately the 
$\bra{\phi_i}$ and $\ket{\psi_i}$ while obeying the biorthogonality 
constraint.  If we have approximate left- and right-eigenvectors 
$\bra{\phi_i'}$ and $\ket{\psi_i'}$ and eigenvalues $\omega_i'$, we can define 
\begin{equation}
W = \sum_i \ket{\psi_i'} \omega_i' \bra{\phi_i'} - H_0.
\label{DefW}
\end{equation}
Then by construction, $H_0 + W$ will have the desired eigenvalues and 
eigenvectors so long as $\braket{\phi_i'}{\psi_j'} = \delta_{ij}$.
We consider several ways to proceed.  

The simplest approach is to build $H_0 + W$ using the eigenvectors of 
$H_0$.  That is, we could define 
\begin{equation}
W_0 = \sum_i \ket{\chi_i} \omega_i \bra{\chi_i} - H_0.
\end{equation}
This assumes that getting the proper broadening and shifts is all that is 
really needed to capture the proper physics.

Our next step is to introduce the dual spaces to $\ket{\psi_i}$ and 
$\bra{\phi_i}$.  That is, we have vectors $\bra{\bar{\psi}_i}$ and 
$\ket{\bar{\phi}_i}$ which satisfy
$\braket{\bar{\psi}_i}{\psi_j} = \delta_{ij}$ 
and 
$\braket{\phi_i}{\bar{\phi}_j} = \delta_{ij}$.
With these in hand, we can now build two approximations to $W$, namely
\begin{subequations}
\begin{eqnarray}
\bar{W}^{\psi} &=& \sum_i \ket{\psi_i} \omega_i \bra{\bar{\psi}_i} - H_0,
\label{dualW_psi}
\\
\bar{W}^{\phi} &=& \sum_i \ket{\bar{\phi}_i} \omega_i \bra{\phi_i} - H_0.
\label{dualW_phi}
\end{eqnarray}
\label{dualW}
\end{subequations}
Both $H_0 + \bar{W}^{\psi}$ and $H_0 + \bar{W}^{\phi}$ have the correct 
eigenvalues; the former gives the correct right-hand eigenvectors but
rotated left-hand eigenvectors, and the latter does the opposite.  Provided we 
are able to define the dual spaces, \eqref{dualW_psi} and \eqref{dualW_phi} 
give different potentials. However, we expect them to yield identical
results as the same physics is carried by either the left- or right- hand
eigenfunctions.

Finally, we consider a further alternative.  We define 
\begin{equation}
\bar{W} = \frac{\bar{W}^{\psi} + \bar{W}^{\phi}}{2},
\end{equation}
which assumes that symmetrizing will have useful effects.  In this case, 
however, neither the eigenvalues nor the eigenvectors will be correct.

Note that, to lowest order in $G_0 \Sigma$, where $G_0$ is the bare Green's 
function (that is, $G_0 = [\epsilon_i - H_0 + \mathrm{i}\eta]^{-1}$), the 
various states we use to build the sundry approximations to $W$ are the same, 
so one expects that all these energy-independent approximations to the 
self-energy would yield roughly similar results.  Also, we have really defined 
the complex potential in Hilbert space, and not as some explicit real-space 
function.  In general, $W$ may be nonlocal and have both real and imaginary 
parts, like $\Sigma(E)$ and unlike the phenomenological complex potentials
in use.

\subsection{Matrix Formulation}
It may prove helpful to put everything in matrix language briefly.  We suppose 
we have a bare Hamiltonian matrix $\H_0$, which has eigenvectors $\mathbf{X}$ 
so that the eigenvalue problem of (\ref{BareEquations}) becomes
\begin{subequations}
\begin{eqnarray}
\H_0 \mathbf{X} &=& \mathbf{X} \bm{\epsilon},
\\
\mathbf{X}^{\dagger} \H_0 &=& \bm{\epsilon} \mathbf{X}^{\dagger},
\\
\mathbf{X}^{\dagger} \mathbf{X} &=& \mathbf{1}.
\end{eqnarray}
\end{subequations}
Once we add the self-energy matrix $\S(E)$, the Dyson equation of 
(\ref{Dyson}) becomes
\begin{subequations}
\begin{eqnarray}
\left[\H_0 + \S(\omega_i) \right] \U_i &=& \omega_i \U_i,
\\
\V_i^{\dagger} \left[\H_0 + \S(\omega_i) \right] &=& \omega_i \V_i^{\dagger} .
\label{Eq4VDagger}
\end{eqnarray}
\end{subequations}
We build up total eigenvector matrices $\U$ and $\V^{\dagger}$ from the 
individual eigenvectors, and similarly build a total eigenvalue matrix 
$\bm{\omega}$.  Note that by taking the transpose of (\ref{Eq4VDagger}), we 
see that if both $\H_0$ and $\S(\omega_i)$ are symmetric matrices, then 
$\V^{\star} = \U$ so that $\V^{\dagger} = \U^{\mathrm{T}}$.

Our various approximations for the complex potential then give us
\begin{subequations}
\begin{eqnarray}
\W_0            &=& \mathbf{X} \bm{\omega} \mathbf{X}^{\dagger} - \H_0,
\\
\bar{\W}^{\psi} &=& \U \bm{\omega} \U^{-1} - \H_0,
\\
\bar{\W}^{\phi} &=& \V^{-\dagger} \bm{\omega} \V^{\dagger} - \H_0,
\\
\bar{\W}        &=& \frac{\bar{\W}^{\psi} + \bar{\W}^{\phi}}{2}.
\end{eqnarray}
\label{ourW}
\end{subequations}
If $\V^{\dagger} = \U^{\mathrm{T}}$, then $\bar{\W}$ is symmetric.

\section{Application}
\label{sec:Application}
In the previous section, we presented a formal derivation of an 
energy-independent CAP from the self-energy that couples the subspace 
Hamiltonian to the continuum states.  In what follows, we test the various 
approximations for the complex potential $W$ and examine its structure in the 
case of transmission of electrons through an atomic chain.

\subsection{Model System}
\label{sec:ModelSystem}
To simplify the calculations and to compare with previous results based on
the more conventional form of CAPs, we turn to a simple H\"uckel model for an
atomic chain with one orbital per atomic site and nearest neighbour 
interactions.  Physically, this is intended as a (very simplistic) treatment 
of an electrode-molecule-electrode system.  Typical molecular junctions are 
made of $\pi$-conjugate carbon chains bonded by anchor groups to metal 
electrodes.

Our model Hamiltonian reads
\begin{equation}
H = H_{\mathrm{L}} + H_{\mathrm{M}} + H_{\mathrm{R}} + V,
\end{equation}
where $H_{\mathrm{L}(\mathrm{R})}$ describes the left (right) electrode, 
$H_{\mathrm{M}}$ describes the molecule, and $V$ describes the 
electrode-molecule coupling.  Using the $c^k_i$ ($(c^k_i)^{\dagger}$) 
operator that creates (annihilates) an electron on the $i^{\textrm{th}}$ site 
of region $k$, the various components of the Hamiltonian are
\begin{subequations}
\begin{eqnarray}
H_{\mathrm{L}} &=& \sum_{i=1}^{\infty} [\varepsilon_L (c^L_i)^{\dagger} c^L_i - \gamma_L (c^L_i)^{\dagger} c^L_{i \pm 1}],
\\
H_{\mathrm{R}} &=& \sum_{i=1}^{\infty} [\varepsilon_R (c^R_i)^{\dagger} c^R_i - \gamma_R (c^L_i)^{\dagger} c^R_{i \pm 1}],
\\
H_{\mathrm{M}} &=& \sum_{i=1}^{N} [\varepsilon_M (c^M_i)^{\dagger} c^M_i - \gamma_M (c^L_i)^{\dagger} c^M_{i \pm 1}],
\\
V              &=& -[(\Gamma_L (c^L_1)^{\dagger} c^M_1 + \Gamma_R (c^M_N)^{\dagger} c^R_1) + \textrm{h.c.}].
\end{eqnarray}
\label{Huckel}
\end{subequations}
Unless stated otherwise, we take 
$\varepsilon_L = \varepsilon_R = \varepsilon_M = \epsilon_0$, 
$\gamma_L = \gamma_R = \gamma_M = \gamma$, and 
$\Gamma_L = \Gamma_R = \gamma/2$.  The same model has been studied using the 
exact self-energy~\cite{GiorgosHuckelCAP1,GiorgosHuckelCAP2}, and has also 
been investigated with a complex potential of the usual 
type~\cite{KopfSaalfrank}.

The Hamiltonian of \eqref{Huckel} has an infinite number of degrees of freedom 
due to the infinite size of the electrodes.  We relate to our previous 
discussion in sections~\ref{sec:Intro} and~\ref{sec:Theory} by considering 
$\NL$ ($\NR$) sites from the left (right) electrode and the $N$ sites 
representing the molecule.  The bare Hamiltonian matrix $\H_0$ of this 
subsystem is thus of dimension $\NT = \NL + N + \NR$, and is
tridiagonal; the diagonal elements are all filled with $\epsilon_0$, and the 
subdiagonal (and superdiagonal) elements are given by $-\gamma$, except for 
the elements that represent the coupling between the molecule and the 
electrode, which are given by $-\Gamma$.

The self-energy needed to account for the rest of the (infinite) electrodes 
({\it i.e.}, to describe the coupling to the continuum) can be calculated 
exactly for this model~\cite{GiorgosHuckelCAP2}.  The expression reads
\begin{equation}
\Sigma(E) = 
  \begin{cases}
      \gamma (\eta - \mathrm{i} \sqrt{1 - \eta^2})   &:  \quad |\eta| \le 1,
   \\
      \gamma (\eta -            \sqrt{\eta^2 - 1})   &:  \quad \text{else},
  \end{cases}
\end{equation}
where 
\begin{equation}
\eta = \frac{E - \epsilon_0}{2 \gamma}.
\label{defEta}
\end{equation}
Note that if we do not explicitly include any of the electrode sites, 
{\it i.e.}, we take $\NL = 0$ or $\NR = 0$, we must scale the self-energy for 
that electrode by $(\Gamma/\gamma)^2$.  

In any event, the total self-energy matrix $\S(E)$ can be written as the sum 
of the self-energy matrices for the left and right electrodes, respectively 
$\SL(E)$ and $\SR(E)$, which are given by
\begin{subequations}
\begin{eqnarray}
\left[\SL(E)\right]_{ij} &=&
	\begin{cases}
	 \Sigma(E) &: \quad i = j = 1,
	\\
	 0         &: \quad \text{else},
	\end{cases}
\\
\left[\SR(E)\right]_{ij} &=&
	\begin{cases}
	 \Sigma(E) &: \quad i = j = \NT,
	\\
	 0         &: \quad \text{else}.
	\end{cases}
\end{eqnarray}
\label{Sigma}
\end{subequations}

\subsection{Results}
\label{sec:Results}
Using the self-energy of \eqref{Sigma}, one can calculate the transmission 
function through the molecule via
\begin{equation}
T(E) = \mathrm{tr}(\LL \, \mathbf{G} \, \LR \, \mathbf{G}^{\dagger}),
\end{equation}
where $\mathbf{G} = [E - (\H_0 + \S(E))]^{-1}$ is the Green's function matrix, 
and $\LL$ and $\LR$ are the spectral densities for the left and right 
electrode, respectively, defined in terms of the self-energies as
\begin{equation}
\bm{\Lambda}_{\mathrm{L(R)}} = 
   \mathrm{i} \left(\S_{\mathrm{L(R)}} - \S_{\mathrm{L(R)}}^{\dagger}\right).
\end{equation}
If one uses a complex absorbing potential instead of the self-energy, one 
simply replaces the self-energy matrices $\SL(E)$ and $\SR(E)$ with complex 
potential matrices $\WL$ and $\WR$ when defining the spectral densities ($\LL$ 
and $\LR$) and the Green's function ($\mathbf{G}$).

We proceed with the analysis of the transmission spectrum for our model.
To this end, a useful measure of the deviations of the transmission $T_W$ using
our approximations to $W$ from the exact result $T_\Sigma$ is provided via

\begin{equation}
\DTM \equiv \max_\eta | T_W(\eta) - T_\Sigma(\eta) |
\label{dev1}
\end{equation}

and

\begin{equation}
\DTA \equiv \frac {1}{2} \int^{1}_{-1} {\mathrm d}\eta \; \;
| T_W(\eta) - T_\Sigma(\eta) | ;
\label{dev2}
\end{equation}
$\eta$ is a dimensionless parameter defined in \eqref{defEta}.

\subsubsection{Tests of Different Approximations to the Complex Potential}
We have several possibilities to consider.  The first thing we should 
point out is that, while with the usual local complex potentials it is trivial 
to define the potential in each electrode, this is not quite as obvious with 
our potentials.  That is, if we follow our procedure as outlined in 
Sec.~\ref{sec:Theory}, building the total complex potential $\W$ directly from 
$\H_0 + \SL + \SR$ , we find that it is not, in general, local, and we cannot 
uniquely define $\WL$ and $\WR$.  For the moment, if we use our procedure to 
build $\H_0 + \W$ in one step, we write
\begin{subequations}
\begin{eqnarray}
\WL &=& \begin{pmatrix}
              \W_{LL} &             \W_{LM} & \frac{1}{2} \W_{LR}  \\
              \W_{ML} & \frac{1}{2} \W_{MM} & \bm{0}               \\
  \frac{1}{2} \W_{RL} &      \bm{0}         & \bm{0}
        \end{pmatrix},
\label{WLeft}
\\
\WR &=& \begin{pmatrix}
              \bm{0}  &             \bm{0}  & \frac{1}{2} \W_{LR}  \\
              \bm{0}  & \frac{1}{2} \W_{MM} &             \W_{MR}  \\
  \frac{1}{2} \W_{RL} &             \W_{RM} &             \W_{RR}
        \end{pmatrix}.
\label{WRight}
\end{eqnarray}
\label{WParts}
\end{subequations}
On the other hand, we could simply use $\H_0 + \S_{\mathrm{L(R)}}$ to build 
$\H_0 + \W_{\mathrm{L(R)}}$, in which case we have a clear and unambiguous way 
of building the potential due to each electrode at the cost of doing twice as 
much work.  We wish to consider all four possibilities for building $\W$ given 
$\S(E)$ as listed in \eqref{ourW}, and for each we wish to see whether it 
suffices to build $\W$ directly and from it obtain $\WL$ and $\WR$ via 
\eqref{WParts} or if we must build $\WL$ and $\WR$ separately from the 
beginning.

In Fig.~\ref{fig:BuildWTot}, we show results from each of our four 
approximations (see Eq.~\ref{ourW}),
wherein we generate $\W$ directly; Fig.~{\ref{fig:BuildWPart}  
is identical except that we build $W_{\mathrm{L(R)}}$ separately through 
$\S_{\mathrm{L(R)}}$.  In all cases, we use $\NL = \NR = 100$ and $N = 12$, 
and compare to the exact result generated by using the self-energy.

\begin{figure}[t]
\includegraphics[width=7.2cm]{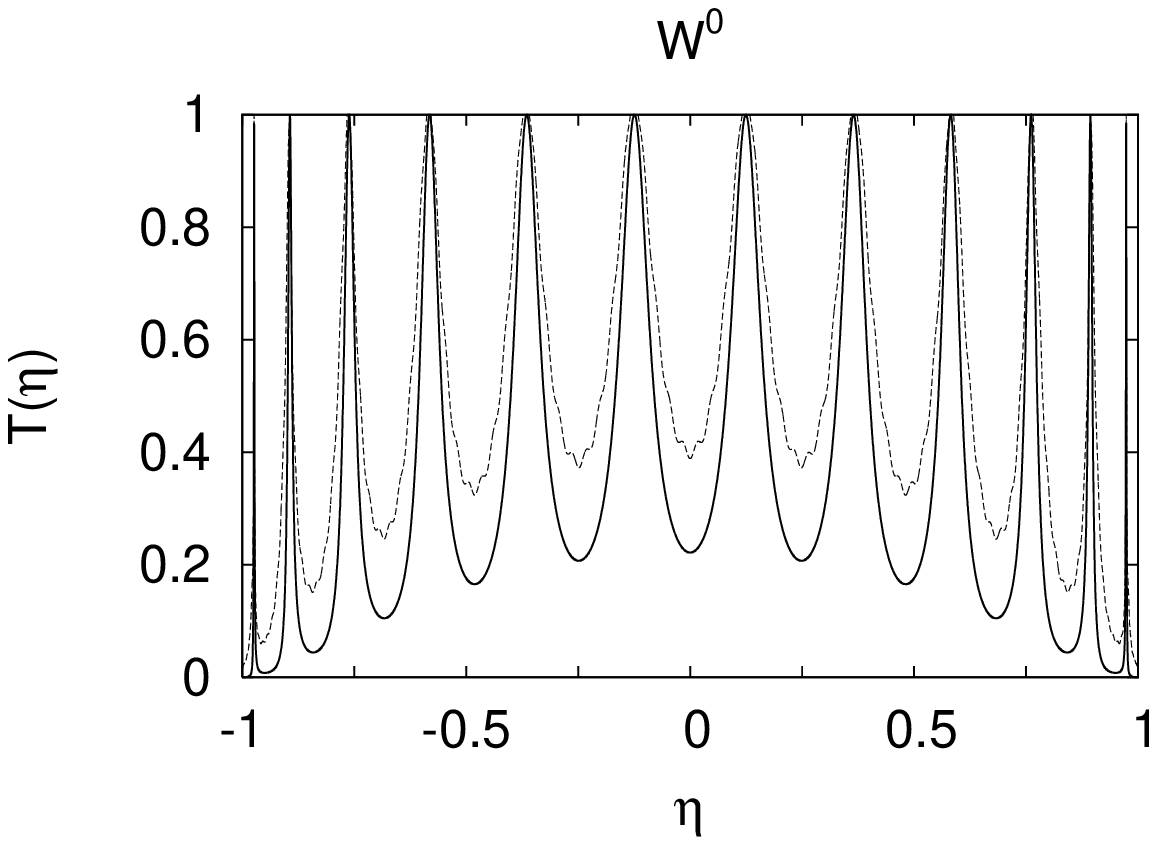}
\\
\vspace{5mm}
\includegraphics[width=7.2cm]{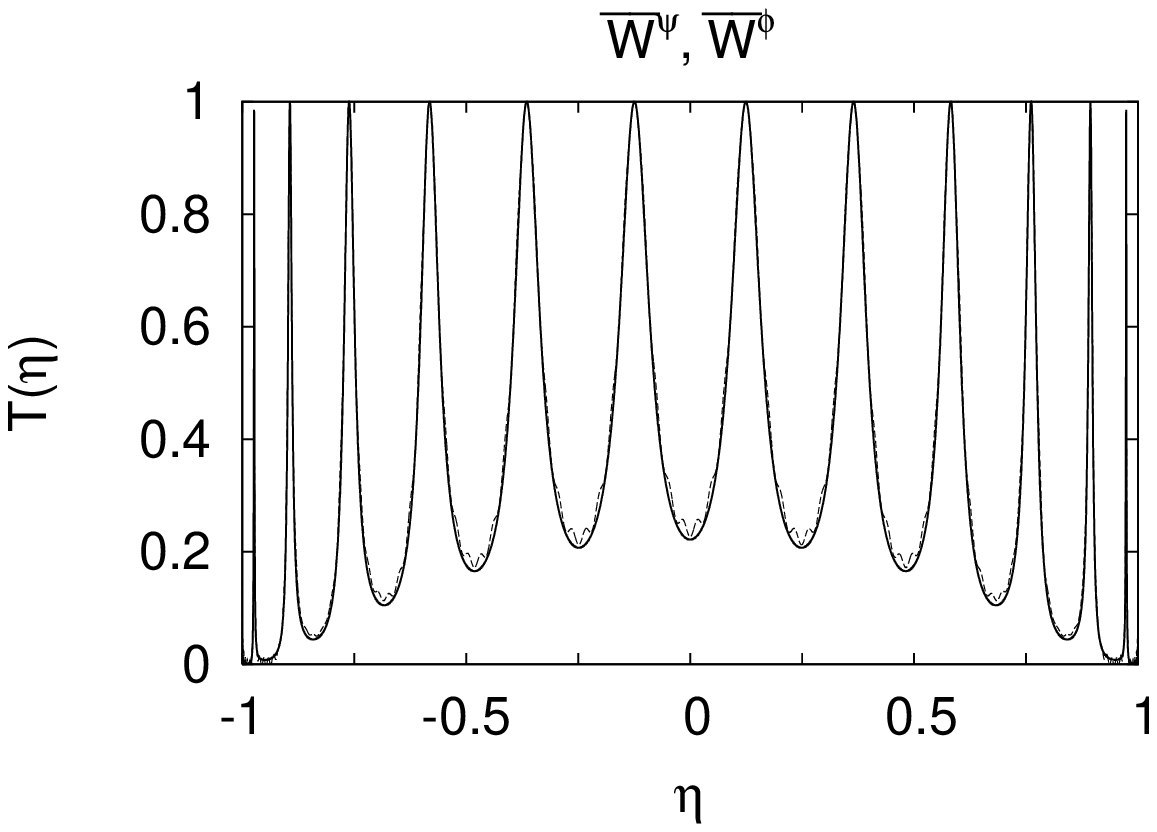}
\\
\vspace{5mm}
\includegraphics[width=7.2cm]{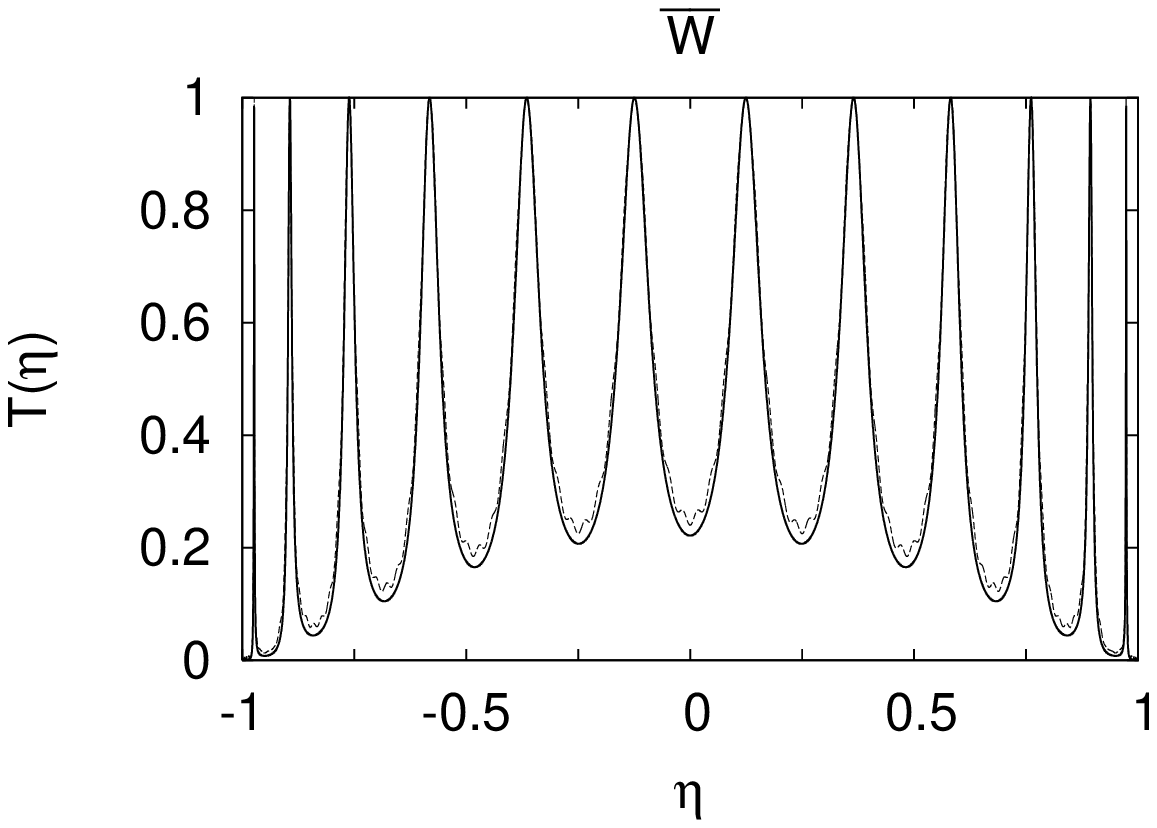}
\caption{Transmission functions with $\NL = \NR = 100$ and $N = 12$ from 
various approximations used to build $\W$ directly.  The exact result is shown 
with the heavy line, while the result from our complex potential is given in 
the dashed line. We recall that $\eta = (E - \epsilon_0) / {2 \gamma}$.
\label{fig:BuildWTot}}
\end{figure}

\begin{figure}[t]
\includegraphics[width=7.2cm]{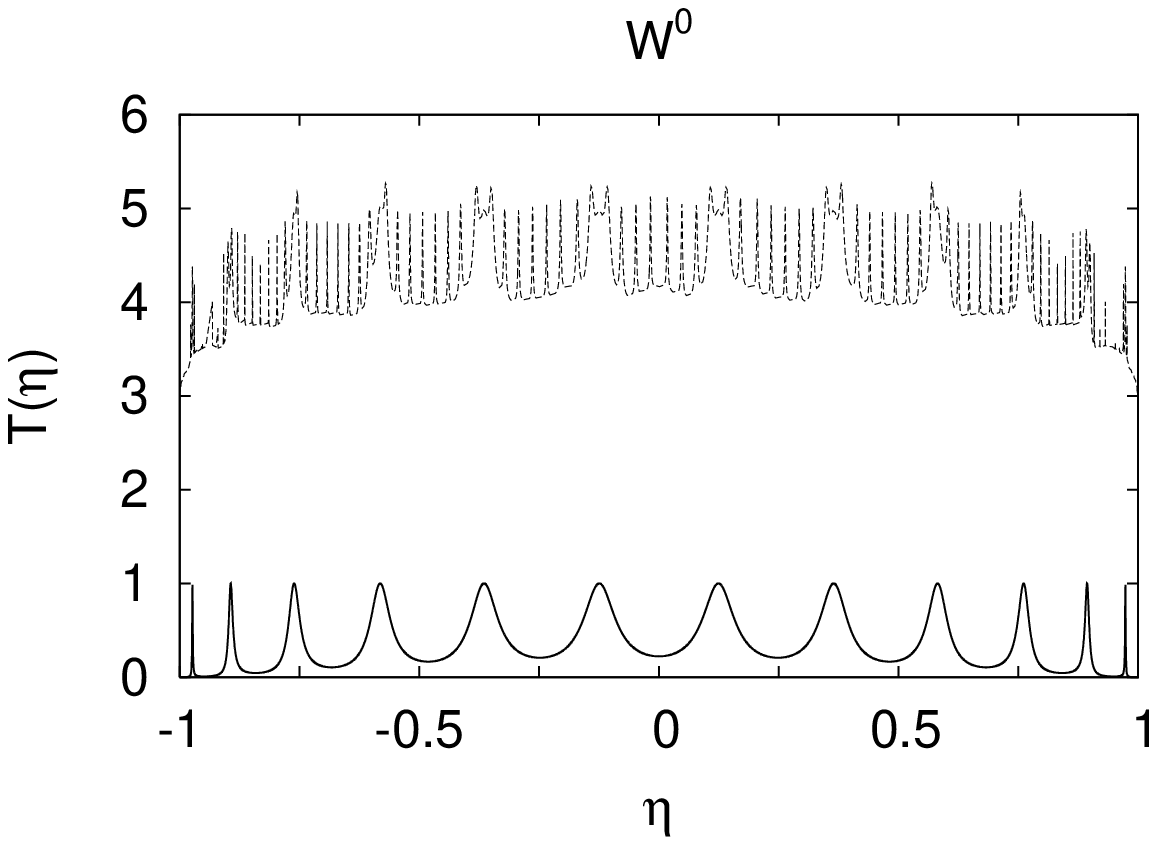}
\\
\vspace{5mm}
\includegraphics[width=7.2cm]{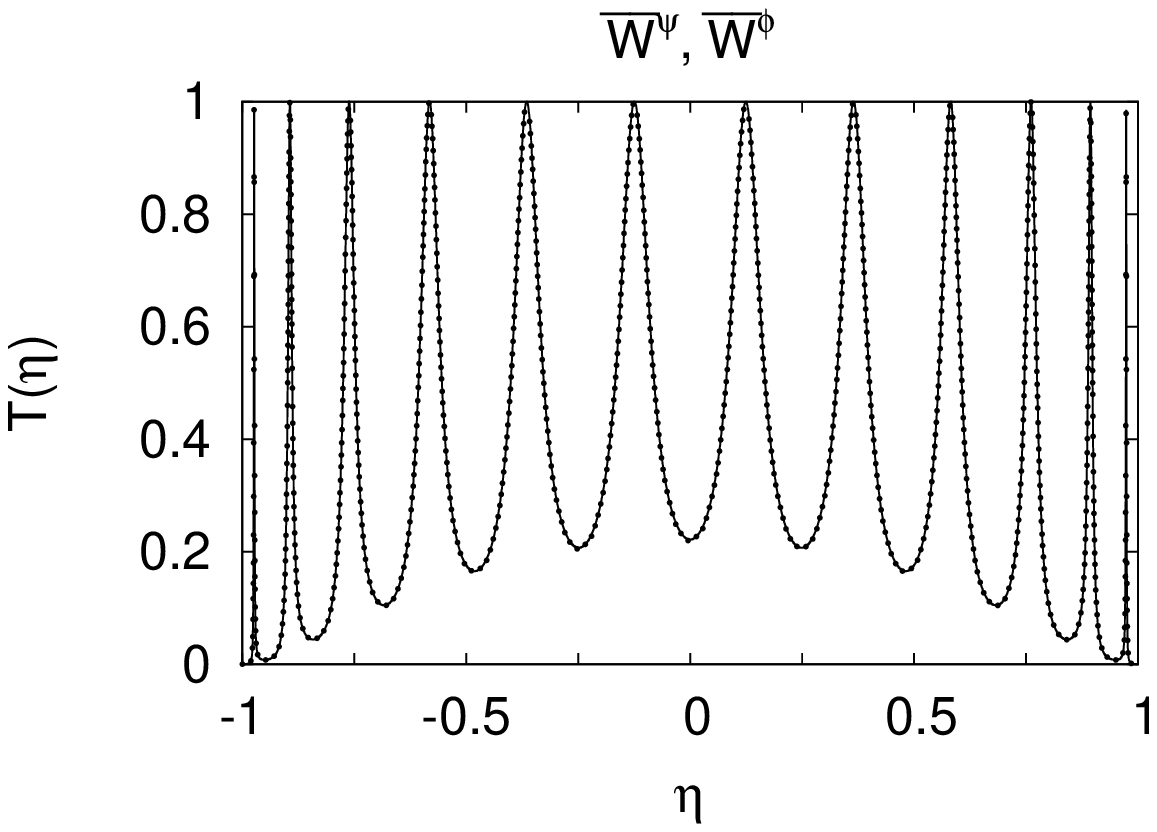}
\\
\vspace{5mm}
\includegraphics[width=7.2cm]{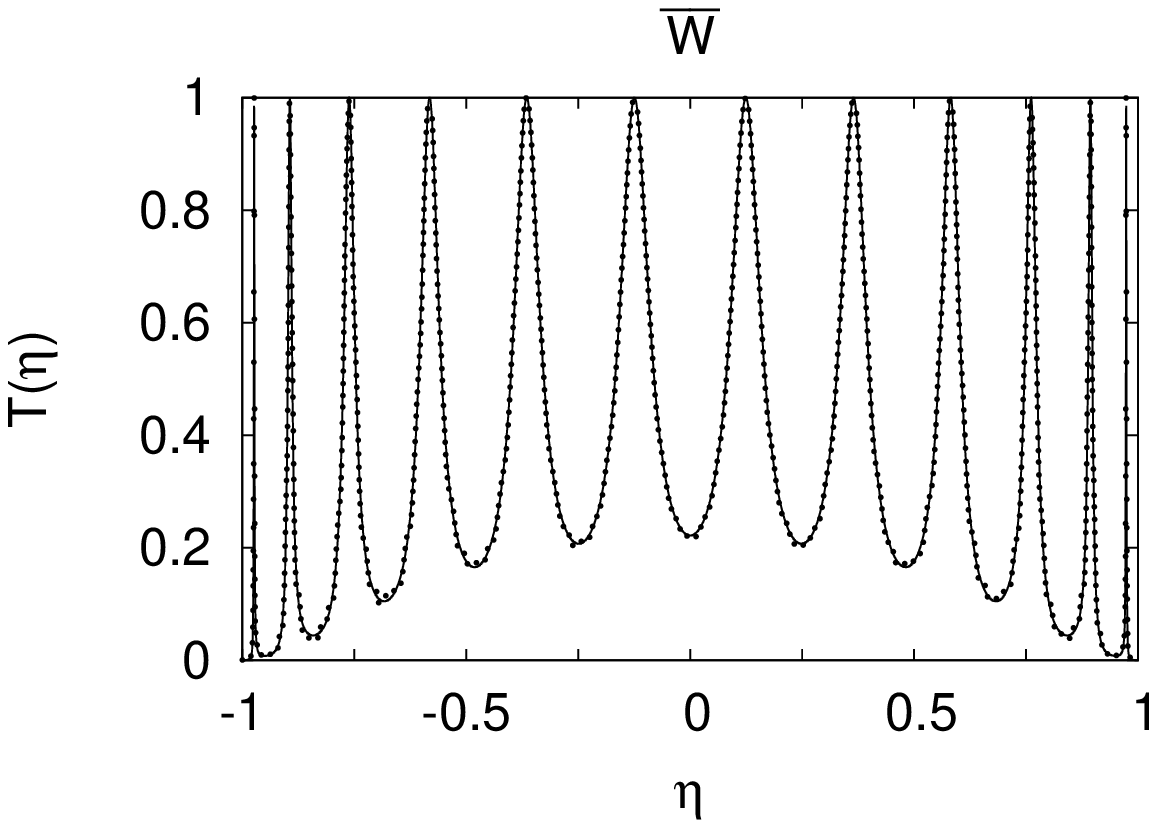}
\caption{Transmission functions with $\NL = \NR = 100$ and $N = 12$ from 
various approximations used to build $\WL$ and $\WR$ separately.
The exact result is shown with the heavy line, while the result from our
complex potential is given in the dashed line (for $W^0$) or in dots (for
$\bar{W}^{\psi}$, $\bar{W}^{\phi}$, and $\bar{W}$).
We recall that $\eta = (E - \epsilon_0) / {2 \gamma}$.
\label{fig:BuildWPart}}
\end{figure}

Let us first examine the results obtained by constructing $\W$ directly and 
extracting $\WL$ and $\WR$ according to \eqref{WParts}.  Although none of 
these results are perfect, we note that most of them are fairly reasonable.  
Using the unperturbed eigenvectors to build $\W$ is clearly inadequate to 
describe the transmission, but already at this crudest level of approximation,
we see that we can obtain some qualitative features.  Unsurprisingly, simply 
using the correct eigenvalues is sufficient to put the resonance peaks in the 
right positions and produce a fair amount of broadening.  The unperturbed 
eigenvectors cannot, however, account for a quantitative description at 
energies far from the resonance peaks, as they yield overestimated 
broadenings.

In order to do a reasonable job over the whole energy range, we 
apparently need to use the correct states.  Using the exact states to build 
$\bar{\W}^{\psi}$ and $\bar{\W}^{\phi}$ gives us a transmission function
that, except for some relatively small oscillations which are most apparent
near the minima, is really quite good. $\bar{\W}^{\psi}$ and $\bar{\W}^{\phi}$
yield identical results, as anticipated, and excluding some effects at the band
edges that we discuss separately, our deviation measures read $\DTM = 0.058$
and $\DTA = 0.014$. Averaging the two, to yield a symmetric complex potential
performs almost as well with $\DTM = 0.071$ and $\DTA = 0.025$.  But
$\H_0 + \bar{\W}$ has neither the correct eigenvalues nor the correct
eigenstates, again pointing to the importance of using the right states.

Though we can already get reasonable transmission functions by building $\W$ 
directly so long as we use the proper states and energy levels, there are 
still noticeable errors which we would like to eliminate.  More important, 
however, is that we have no {\it a priori} way of defining the potential due 
to each electrode separately, and our use of (\ref{WParts}) to separate $\WL$ 
from $\WR$ is largely arbitrary.  We can solve both of these problems by 
building $\WL$ and $\WR$ separately and combining them to build $\W$ at the 
end.

Note first that we must exercise some care if we follow this course.  If we 
were to simply use the eigenvectors $\mathbf{X}$ of the unperturbed 
Hamiltonian to do this, we would obtain $\WL = \WR = 1/2\, \W$, which 
is clearly nonsensical and is why the transmission function built from $\W^0$ 
is meaningless.  To amplify on this point, note that under this construction, 
the spectral densities $\LL$ and $\LR$ are diagonal in the same basis as is 
the Green's function.  If the eigenvalues of $\H_0 + \WL$ and $\H_0 + \WR$ are 
$\omega_i = \epsilon_i + 1/2 \; \delta_i - 1/2 \; \mathrm{i} \, \lambda_i$, 
then in the diagonal basis we have~\cite{note2}
\begin{subequations}
\begin{eqnarray}
(\LL)_{i,i}                  &=& \lambda_i
\\
(\LR)_{i,i}                  &=& \lambda_i,
\\
(\mathbf{G})_{i,i}           &=& \frac{1}{E - \epsilon_i - \delta_i - \mathrm{i} \lambda_i},
\\
(\mathbf{G}^{\dagger})_{i,i} &=& \frac{1}{E - \epsilon_i - \delta_i + \mathrm{i} \lambda_i}.
\end{eqnarray}
\end{subequations}
The transmission function is thus
\begin{equation}
T(E) = \sum_i \frac{ \lambda_i^2}{(E - \epsilon_i - \delta_i)^2 + \lambda_i^2}.
\end{equation}
In other words, it is the sum of $\NT$ independent Lorentzian resonances; this 
explains the jagged nature of the calculated transmission function, as well as 
the fact that $T(E)$ is far too large.  This should be contrasted with the 
exact result and with our results using other constructions of $\W$, in which 
interference between states suppresses all but $N$ peaks in the transmission 
function.

When, however, we use $\bar{\W}^{\psi}$ or $\bar{\W}^{\phi}$, the results are 
for all practical purposes perfect; $\DTM = 0.006$ and $\DTA = 0.001$.
While we see a slight degradation in quality from using $\bar{\W}$ instead
(with $\DTM = 0.093$ and $\DTA = 0.009$), any of these three approaches would
be reasonable to take. Once again, the importance of generating the right
states is clear.

In the remainder of this work, we will use only $\bar{\W}^{\psi}$, built by 
constructing $\WL$ and $\WR$ separately, as the results clearly indicate that 
this is the best way to construct a complex potential among 
all those that we have considered as demonstrated in the middle panel
of Fig.~\ref{fig:BuildWPart}~\cite{note3}.

\subsubsection{Effects of Electrode Size}
One of the most interesting features of our complex potential is that the 
results for the transmission function show an unexpected insensitivity to 
$\NL$ and $\NR$.  In Fig.~\ref{fig:ElectrodeSizeTest12}, we show results for 
$N = 12$ and for varying electrode sizes $\NL$ and $\NR$.  The results are 
virtually indistinguishable; we can even choose to use $\NL = \NR = 0$ and 
still obtain a remarkably accurate transmission function.  Of course, the 
exact transmission function is strictly independent of these parameters, but 
this observation points to the redundancy of an absorption grid outside the
area of interest in similar applications.  With a complex potential of the
more usual form, this is not necessarily the case.  In common implementations,
a finite absorption region, whose extent depends on the precise form of the
CAP, is essential for the damping of the resonant wave function.

\begin{figure*}
\includegraphics[width=7.2cm]{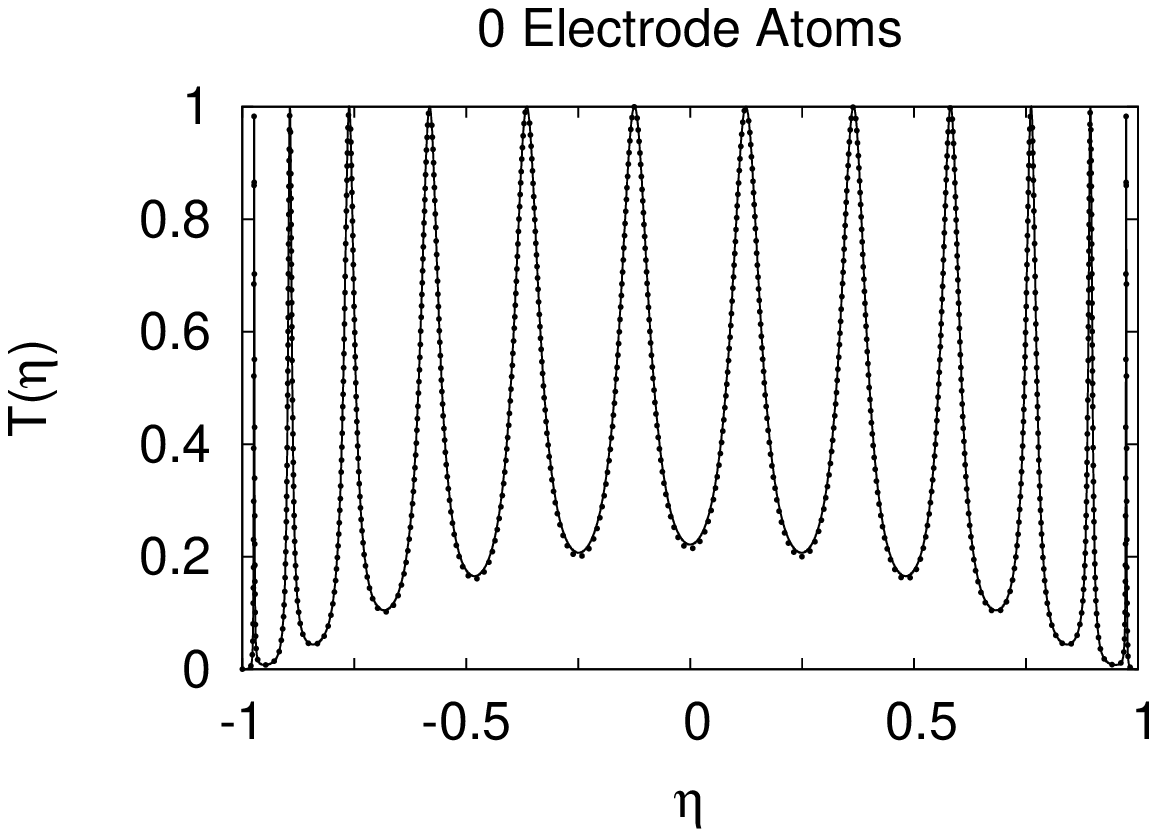}
\hspace{1cm}
\includegraphics[width=7.2cm]{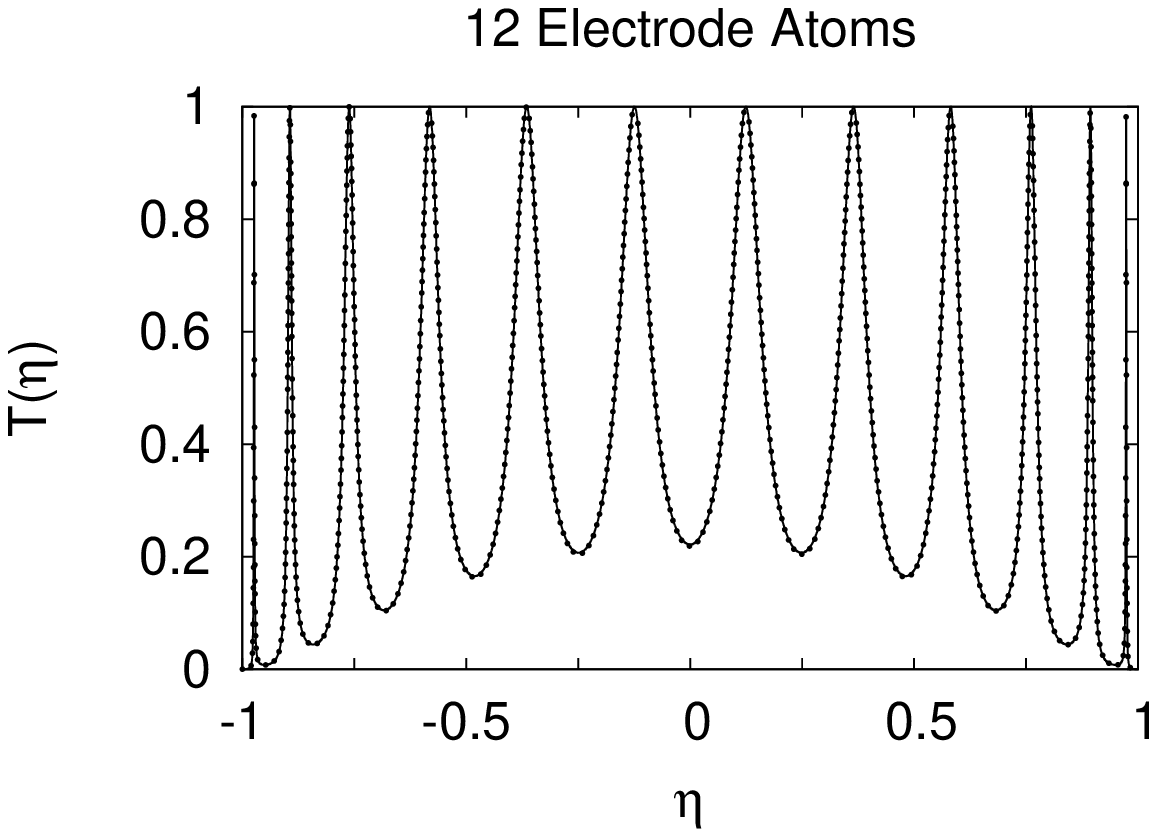}
\caption{Transmission as a function of $\eta = (E - \epsilon_0) / {2 \gamma}$
for $N = 12$ with varying $\NL$ and $\NR$.
The exact result is given by the solid line; our results are the dots.
\label{fig:ElectrodeSizeTest12}}
\end{figure*}

To some extent, this success is illusory.  Even our optimal complex potentials
do not work as well far from resonance or close to the band edges as they do 
near a resonance; any residual deviations from the exact results occur at 
these energies.  While the small errors observed may not be of practical 
relevance, the electrode extent still plays the major role if stringent 
convergence is required.  We demonstrate this with the following examples.

{\it Resonance Minima}
- A careful perusal of Fig.~\ref{fig:ElectrodeSizeTest12} reveals that our 
results are not quite as good near the minima of the transmission 
function as they are in the resonance peaks.  This effect can be seen more 
readily by decreasing $N$, which decreases the number of resonances and 
increases the range of energies far from any of the resonances.  In 
Fig.~\ref{fig:ElectrodeSizeTest3}, we show results for $N = 3$ and varying 
$\NL$ and $\NR$.  We can now see that while once again we can use 
$\NL = \NR = 0$ to describe the transmission function at resonance, we must 
use rather large leads if we are to obtain accurate results far from any of 
the peaks.  Indeed, even with $\NL = \NR = 60$, there are still some residual 
oscillations about the exact result. On the other hand, the deviation measures
\eqref{dev1} and \eqref{dev2} decrease monotonically from $\DTM = 0.037$ and
$\DTA = 0.019$ to $\DTM = 0.016$ and $\DTA = 0.004$ as the electrode size
varies from $\NL = \NR = 0$ to $\NL = \NR = 60$.

\begin{figure*}
\includegraphics[width=7.2cm]{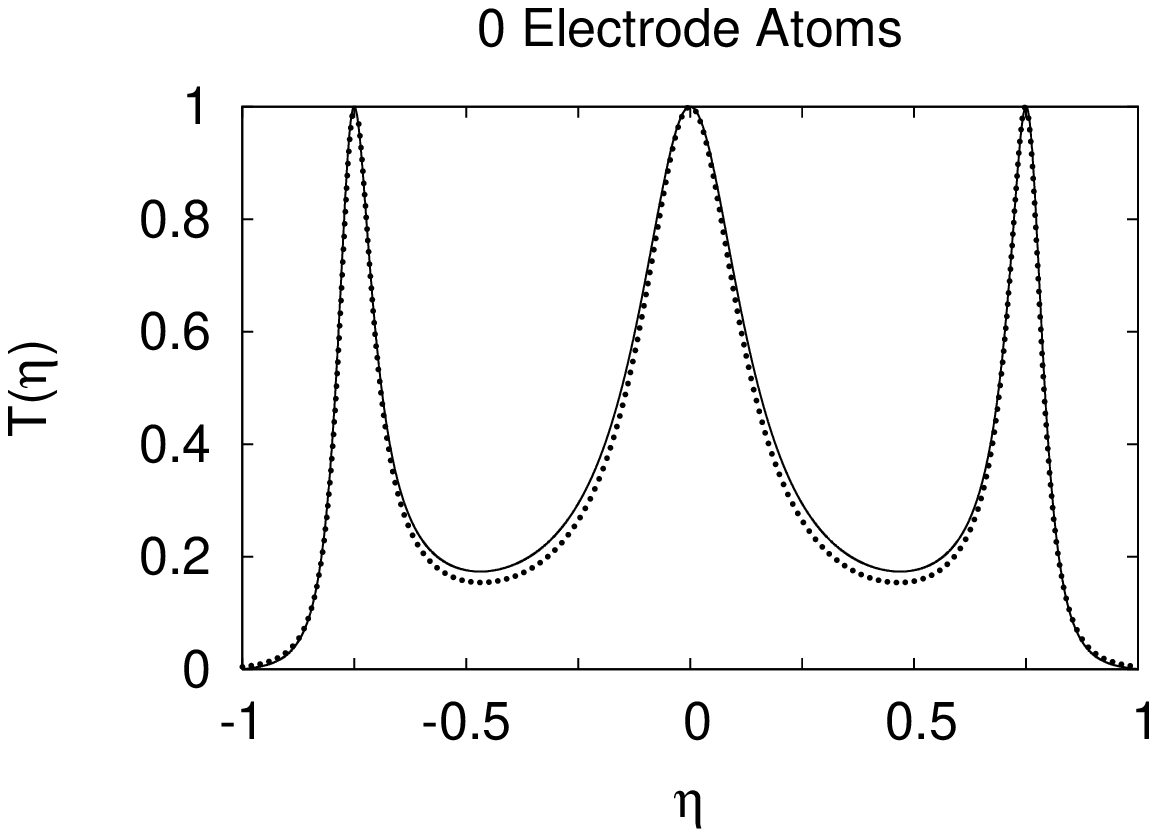}
\hspace{1cm}
\includegraphics[width=7.2cm]{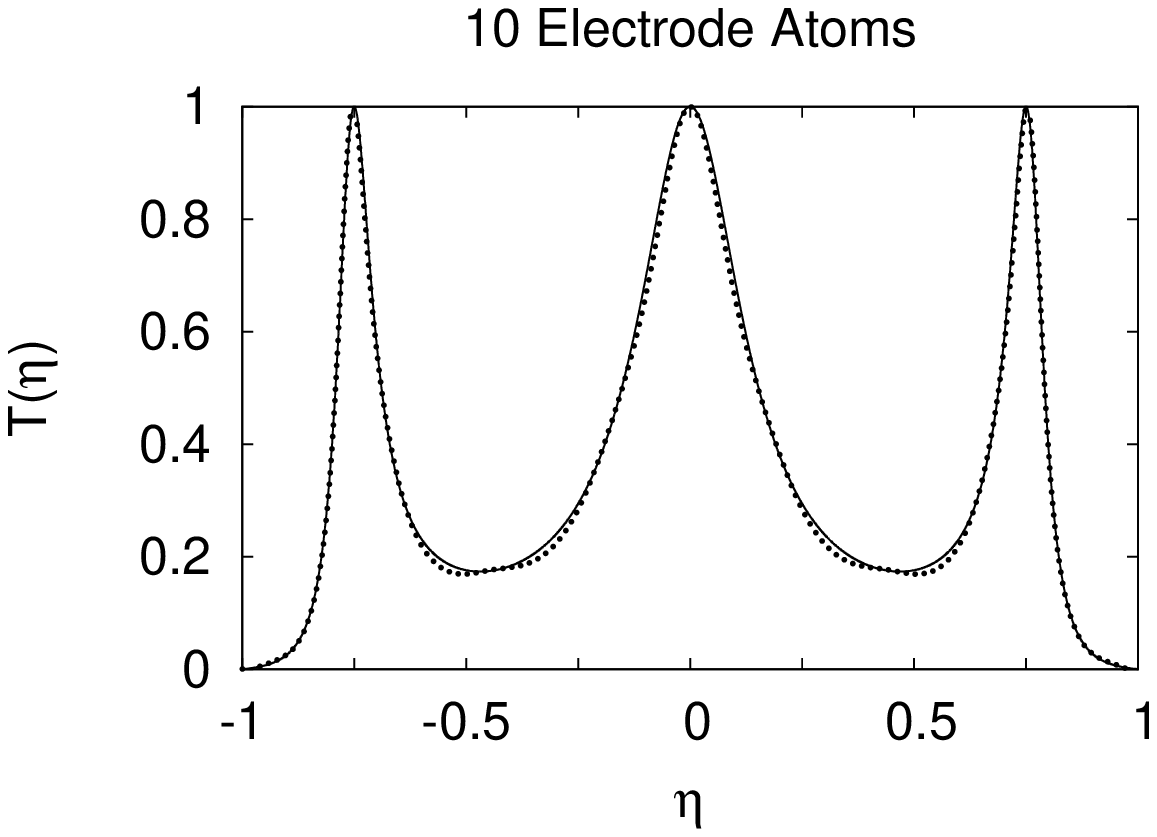}
\\
\vspace{5mm}
\includegraphics[width=7.2cm]{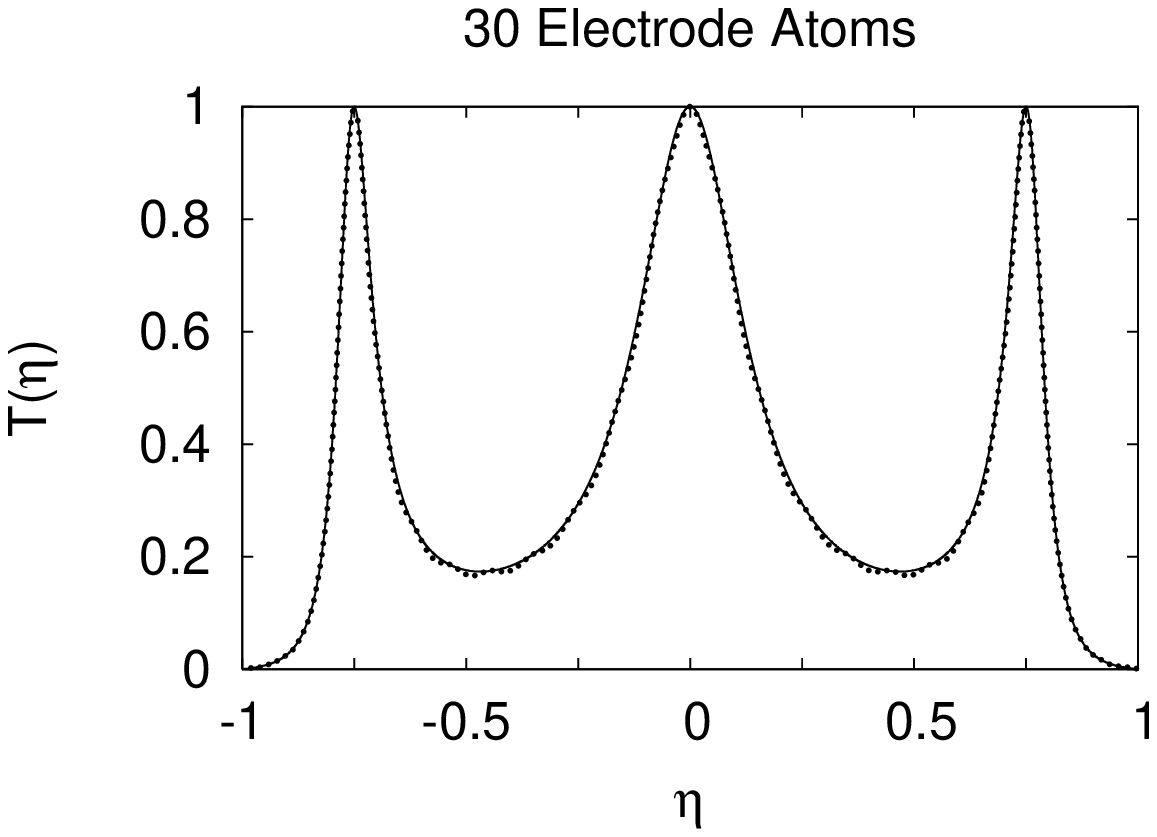}
\hspace{1cm}
\includegraphics[width=7.2cm]{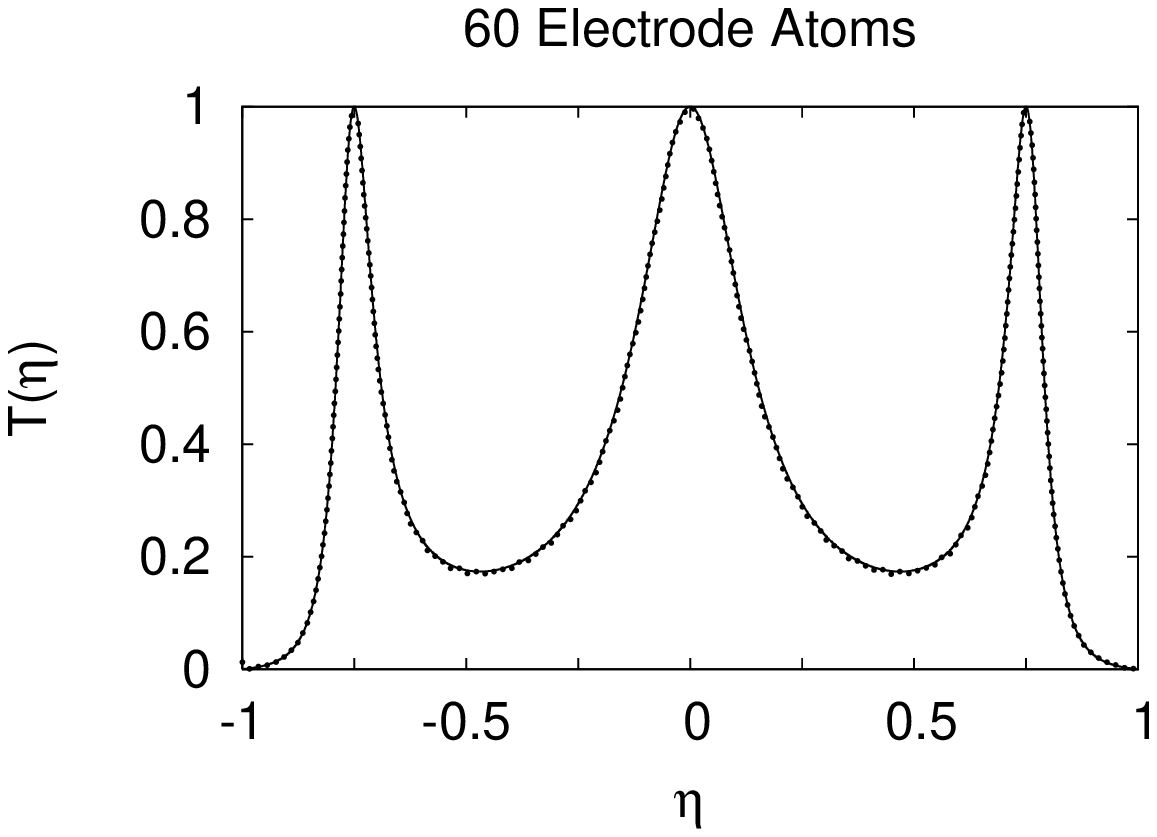}
\caption{Transmission as a function of $\eta = (E - \epsilon_0) / {2 \gamma}$
for $N = 3$ with varying $\NL$ and $\NR$.
The exact result is given by the solid line; our results are the dots.
\label{fig:ElectrodeSizeTest3}}
\end{figure*}

{\it Band Edges}
- One way to see the difficulty in describing the transmission function near 
the band edge is to consider the eigenvalues of the matrix $\mathbf{T}(E)$
given by
\begin{equation}
\mathbf{T}(E) = \LL \mathbf{G} \LR \mathbf{G}^{\dagger},
\end{equation}
whose trace yields the transmission function.
The exact matrix derived from the self-energy has exactly one non-zero 
eigenvalue at each energy; this eigenvalue is of course numerically equal to 
the transmission function at that energy.  This is to be compared with the 
results from our calculations using $\bar{\W}^{\psi}$, displayed in 
Fig.~\ref{fig:TEvalsTest}.  Near the band edges, we have two non-zero 
eigenvalues at each energy, and their sum gives us the calculated transmission 
function.  As the energy moves away from the band edge, the smaller of the two 
eigenvalues goes to zero and we approach the exact result.  The precise energy 
range that marks convergence within a given tolerance depends on the size of 
the included electrodes.  This is readily seen by the eigenvalue plots in 
Fig.~\ref{fig:TEvalsTest} as $\NL$ and $\NR$ increase.

\begin{figure*}
\includegraphics[width=7.2cm]{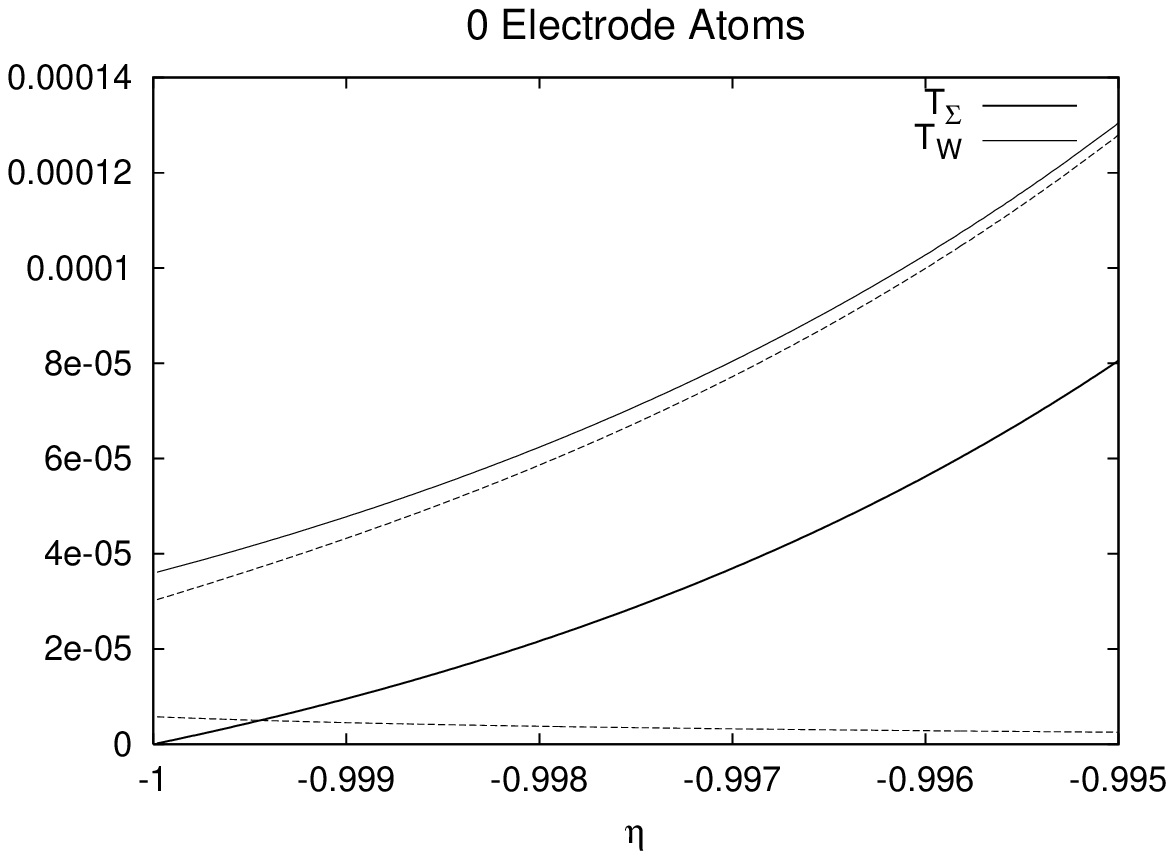}
\hspace{1cm}
\includegraphics[width=7.2cm]{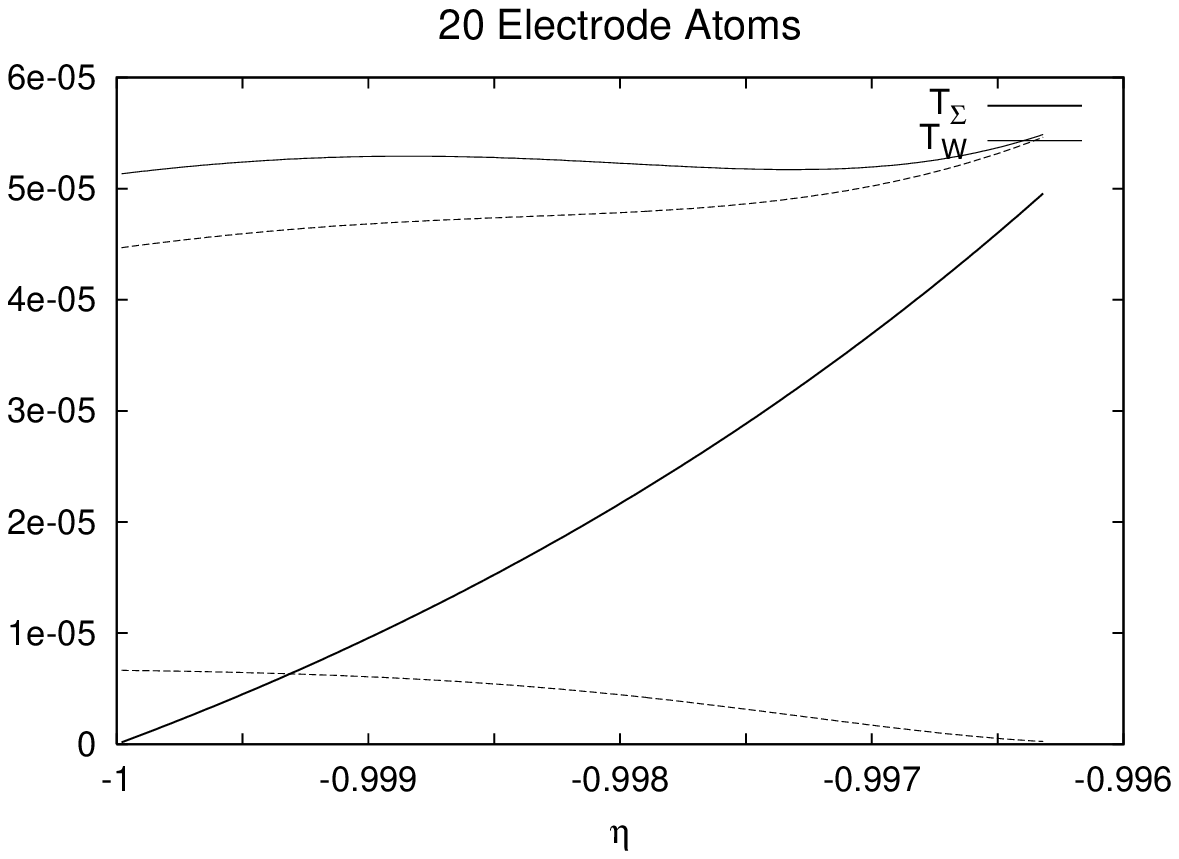}
\caption{Transmission function and non-zero eigenvalues of $\mathbf{T}(\eta)$
for $N = 12$ and varying $\NL$ and $\NR$. The exact transmission function 
($T_{\Sigma}$) is displayed in the heavy line; the transmission function
calculated using the complex potential ($T_W$) is displayed in the light line, 
and the two non-zero eigenvalues of $\mathbf{T}(\eta)$ are given by the dotted 
line. We recall that $\eta = (E - \epsilon_0) / {2 \gamma}$.
\label{fig:TEvalsTest}}
\end{figure*}

{\it Ideal Wire}
- A stricter test of our complex potential can be done by calculating the 
transmission function when the electrode-molecule coupling parameter $\Gamma$
is set equal to the other coupling parameter, $\gamma$.  In this case, the
exact result is that $T(E) = 1$, as it should be for an ideal wire with no 
potential sources for scattering.  We expect that our complex potential would 
find this case more difficult to describe, and that the sensitivity to $\NL$ 
and $\NR$ (or, as we cannot distinguish molecule sites from electrode sites, 
the sensitivity to $\NT$) should be much larger than in our previously 
considered cases.  We show results for $\NT = 72$ and for $\NT = 212$ in 
Fig.~\ref{fig:FlatnessTest}.  Note the reduced range of the ordinate in these 
plots.  In either case, we see small oscillations around the exact result, and 
we note that the results are considerably worse near the band edge than they 
are near the middle of the band.  The results improve by increasing $\NT$, 
again pointing to the role of the electrode size.  We believe that the 
problems near the band edge might be related to the van Hove divergence in the 
electrode density of states at $E = \pm 2 \gamma$.  It is worth noting that 
these same difficulties in describing transmission through an ideal wire are 
found when phenomenological complex potentials are used in the 
calculation~\cite{flat_CAP}.

\begin{figure*}
\includegraphics[width=7.2cm]{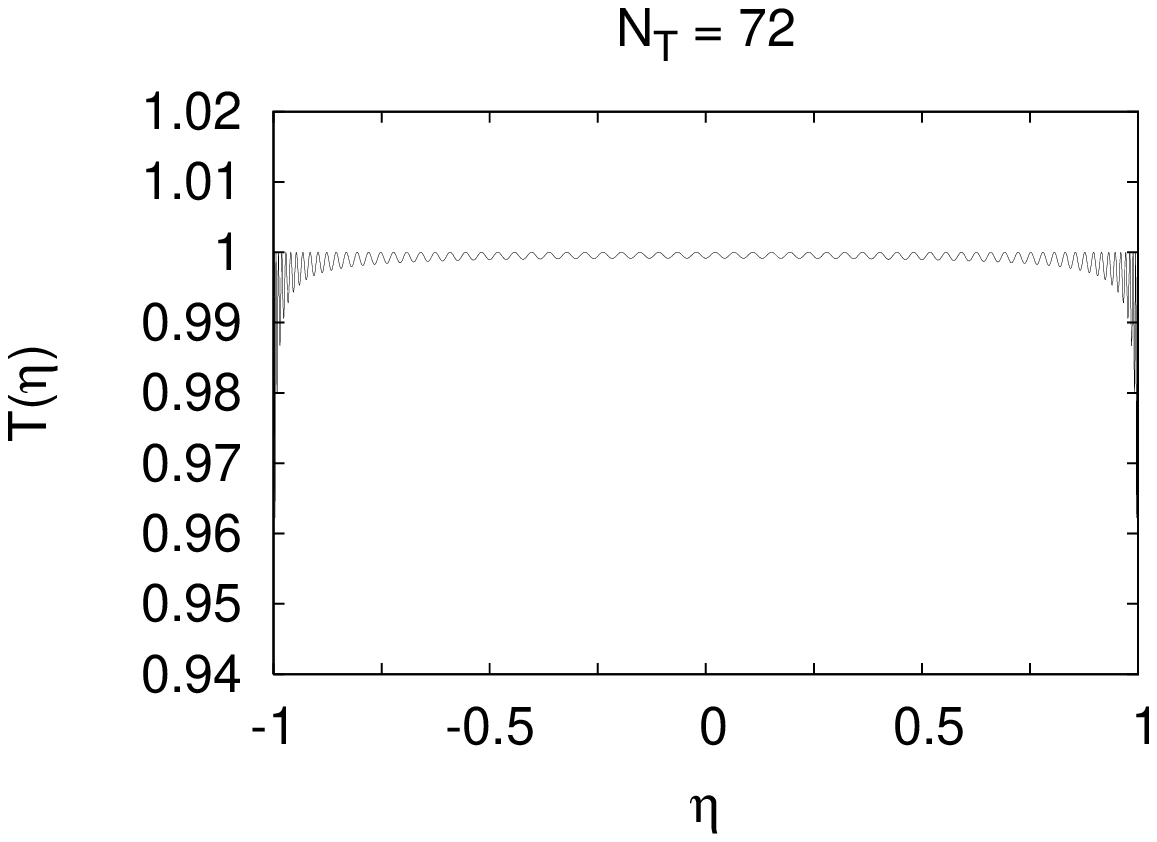}
\hspace{1cm}
\includegraphics[width=7.2cm]{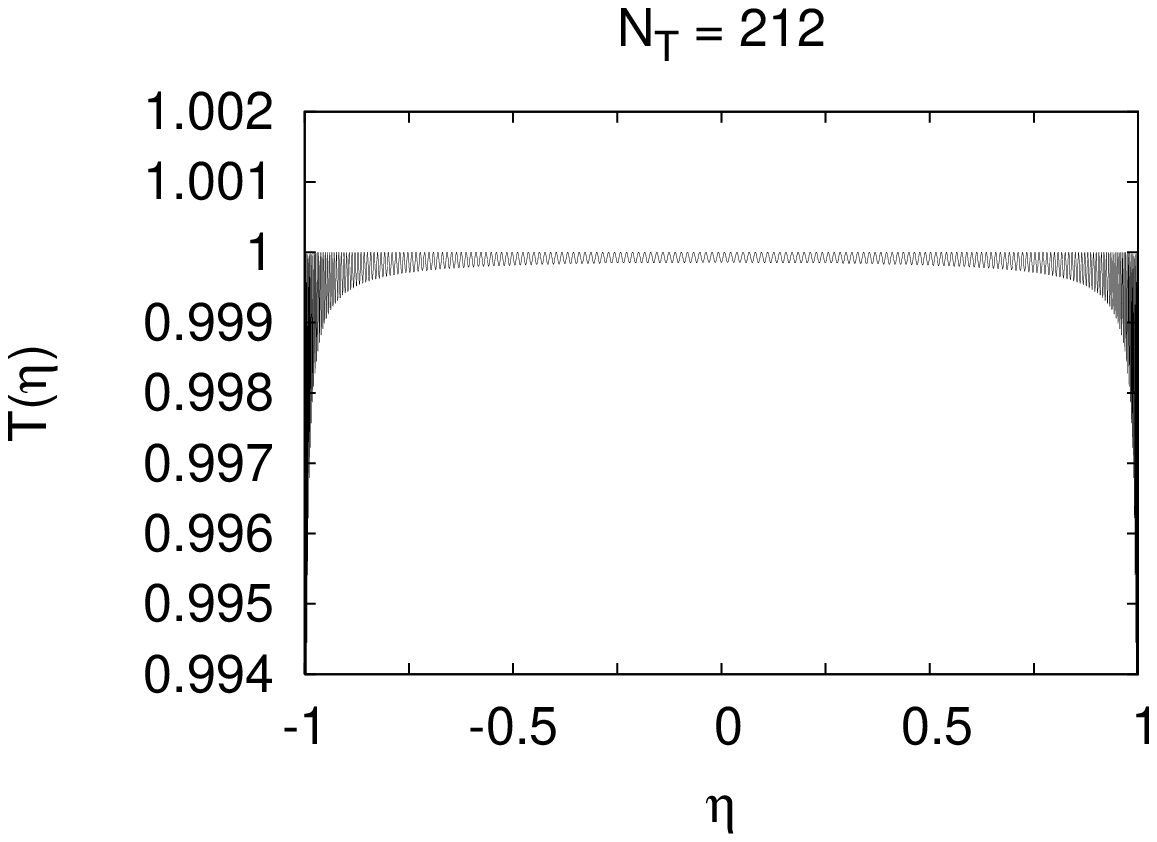}
\caption{Transmission as a function of $\eta = (E - \epsilon_0) / {2 \gamma}$
with $\Gamma = \gamma$ for two different values of $\NT$.
The exact result is that $T(\eta) = 1$.  Notice the range of the ordinate.
\label{fig:FlatnessTest}}
\end{figure*}

\subsubsection{Structure of the Complex Potential: Non-locality}
Now that we have seen how well our complex potential works, it is worthwhile 
investigating its structure.  We will continue to focus on $\bar{\W}^{\psi}$, 
and on $\WL$ and $\WR$ built separately.

With the bare Hamiltonian and the self-energy that we are 
using, $\WL$ and $\WR$ take a particularly simple structure.  For $\WL$, only 
the first row is non-zero, and for $\WR$ only the last row is non-zero; for 
symmetry reasons it is of course the case that $\WR$ can be obtained from 
$\WL$ by left-right reflection (that is, reflection about the skew diagonal).  
Elements of the first row of $\WL$ for $\NL = \NR = 30$ and $N = 12$ are shown 
in Fig~\ref{fig:CAPElements}.  Note that they show a repeating pattern in 
which each imaginary element is followed by a real element, and each real 
element is followed by an imaginary element.  Further, while the magnitude of 
the element typically decreases as one moves from left to right, this is not 
always the case and there are some relatively significant elements coupling 
one end of the left electrode to the other end of the right electrode.  Our 
complex potential, in other words, is strongly non-local.

\begin{figure}[h]
\includegraphics[width=7.2cm]{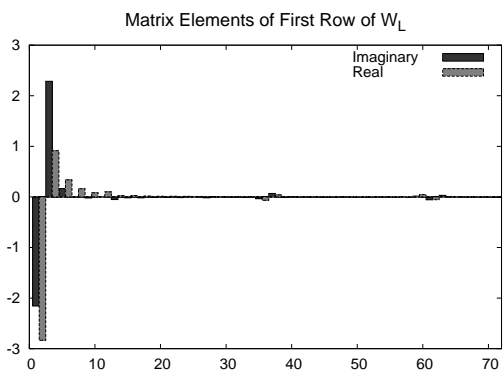}
\caption{Elements of the first row of $\WL$ for $\NL = \NR = 30$ and 
$N = 12$.  Some elements are real; the others are imaginary.  None have both 
real and imaginary parts.
\label{fig:CAPElements}}
\end{figure}

The structure of our complex potential is in sharp distinction to the more 
common complex potentials.  For example, the potential used by Kopf and 
Saalfrank\cite{KopfSaalfrank} has the usual structure, namely,
the only non-zero elements are negative imaginary and on the diagonal.
They vanish within the molecule and the opposite electrode, and increase 
in magnitude as one moves away from the molecular region.

We have seen that our best complex potentials are non-local, but that the 
elements far from the diagonal are rather small.  This naturally raises the 
question of just how important the non-locality is.  We can study this issue 
by simply constructing the full, non-local complex potential, but, once it is 
built, removing all but the diagonal ({\it i.e.}, the fully local part) and 
some number $N_S$ of the subdiagonals.  As $N_S$ approaches $\NT-1$, then, we 
approach the fully non-local CAP in a smooth way.  In 
Fig.~\ref{fig:LocalityTest}, we display results for the transmission function 
with $\NL = \NR = 30$, $N = 12$, and several values of $N_S$.  

Clearly, the local approximation is wholly inadequate, and while including 
only a few subdiagonals improves the results dramatically, there are still 
deviations from the exact result even with $N_S = 30$.  Since there are
no such deviations for the fully non-local complex potential, it is clear that 
the strong non-locality plays a critical role in mimicking the effects of the 
self-energy.

\begin{figure*}
\includegraphics[width=7.2cm]{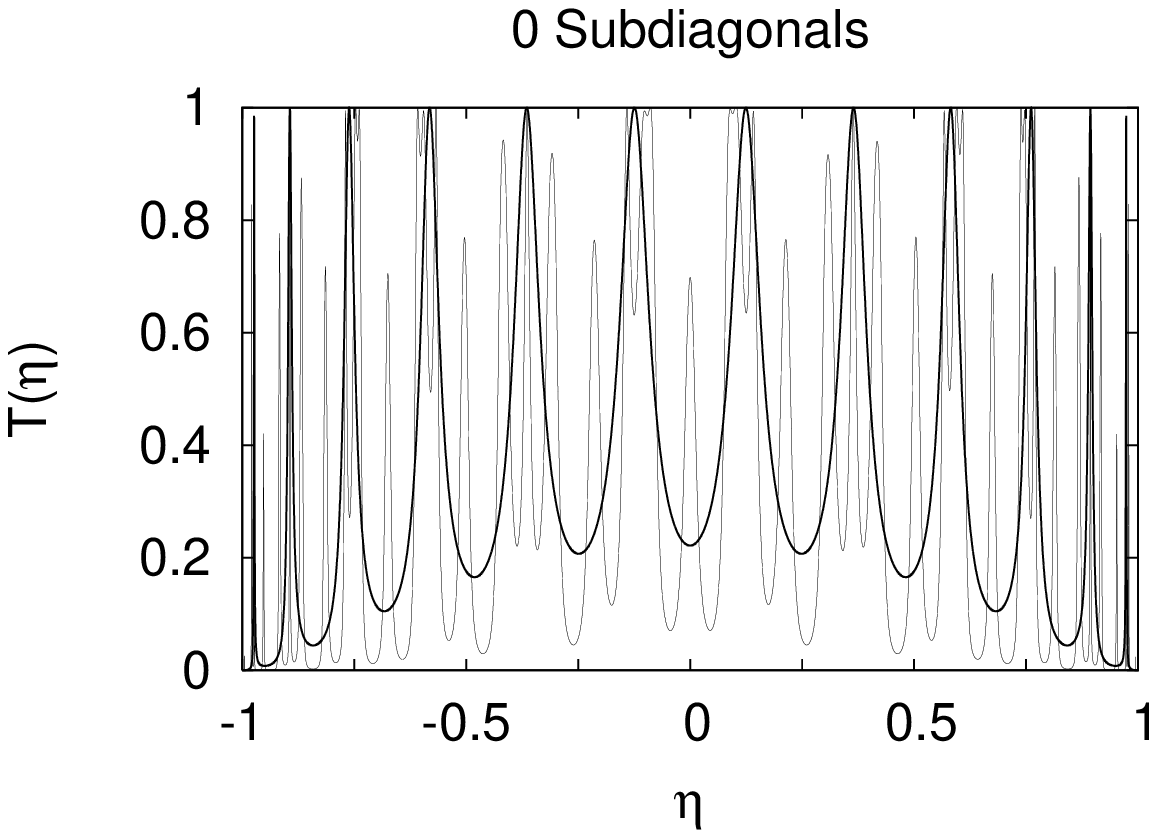}
\hspace{1cm}
\includegraphics[width=7.2cm]{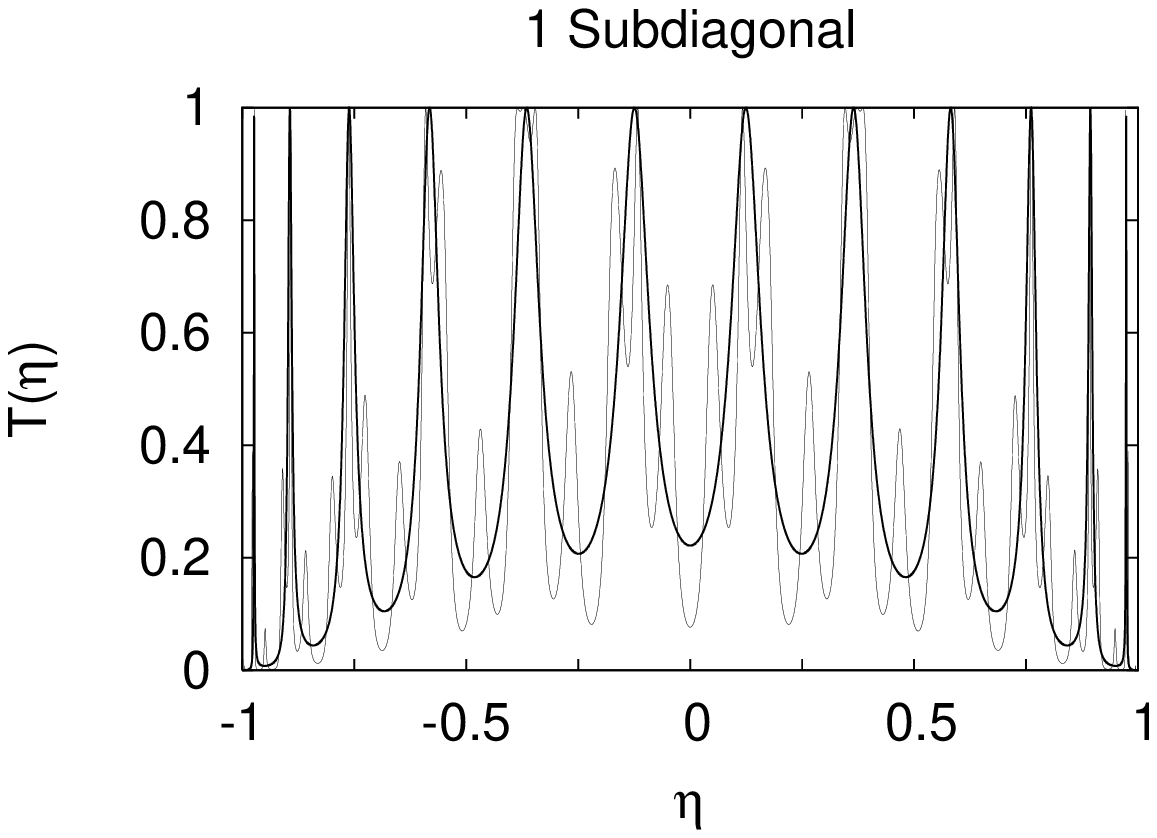}
\\
\vspace{5mm}
\includegraphics[width=7.2cm]{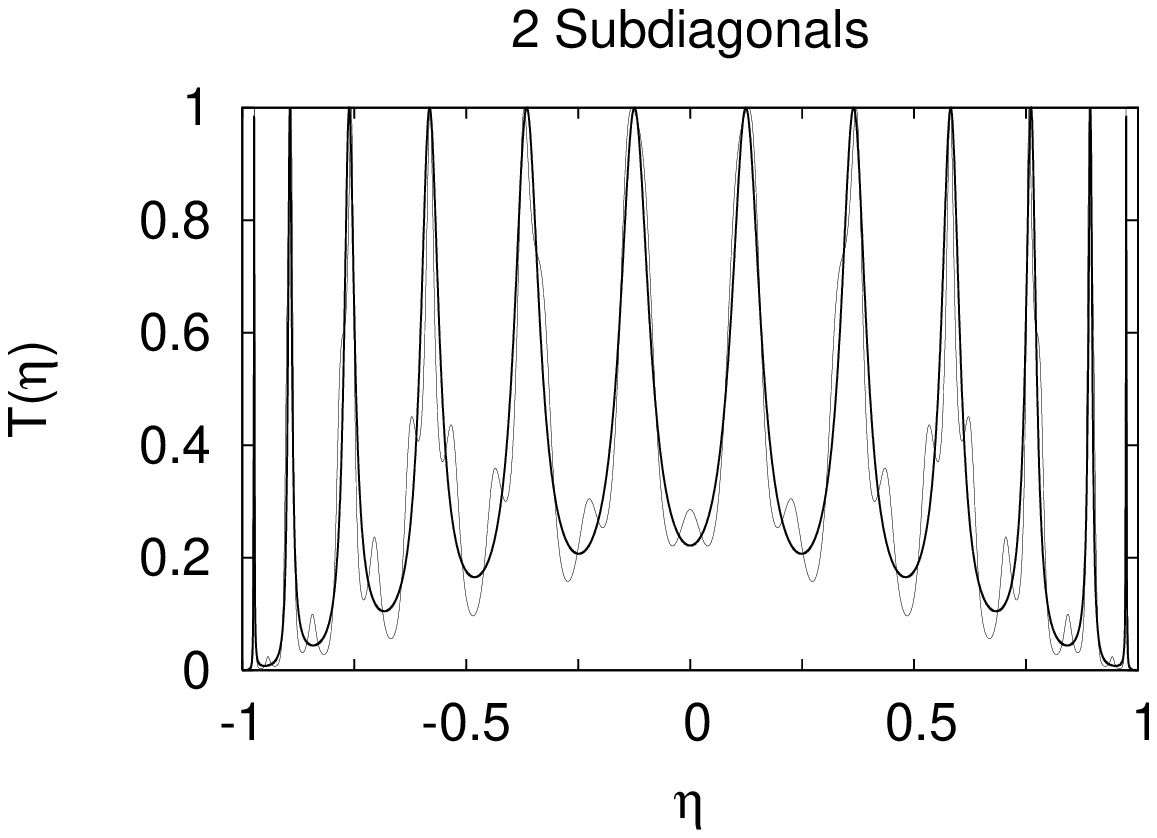}
\hspace{1cm}
\includegraphics[width=7.2cm]{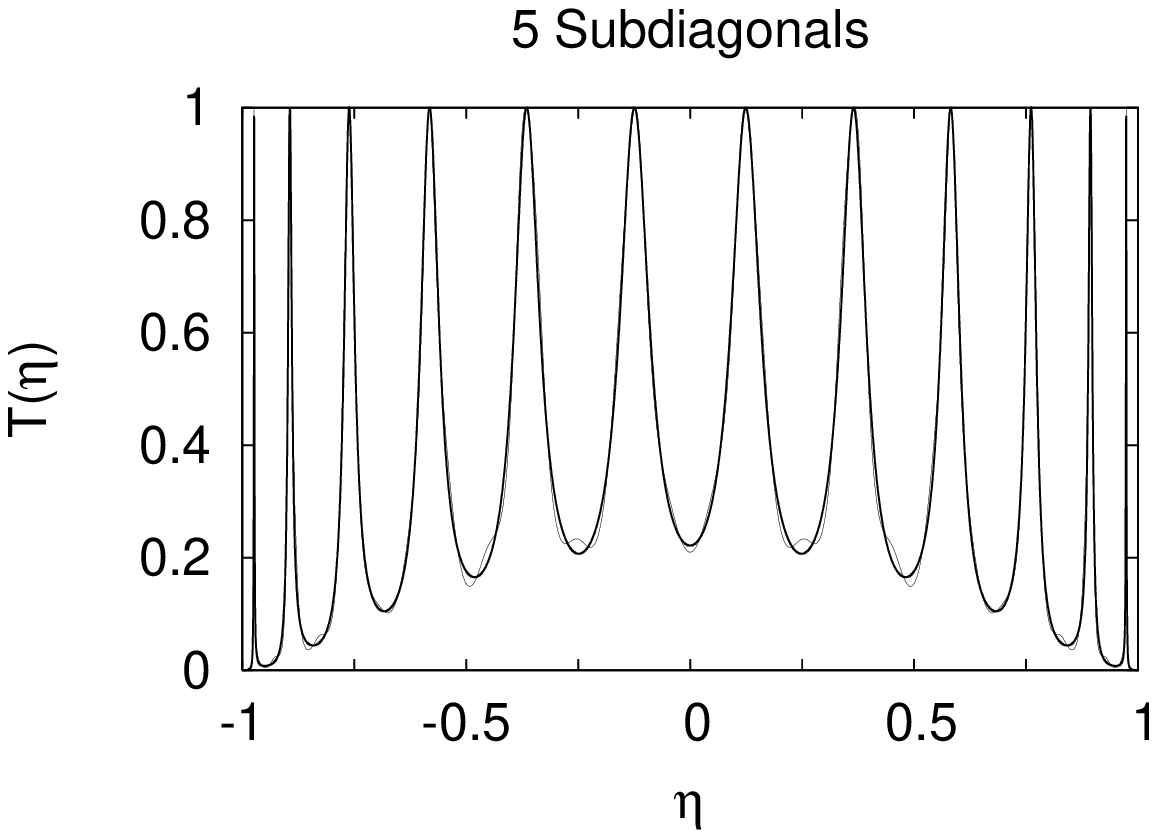}
\\
\vspace{5mm}
\includegraphics[width=7.2cm]{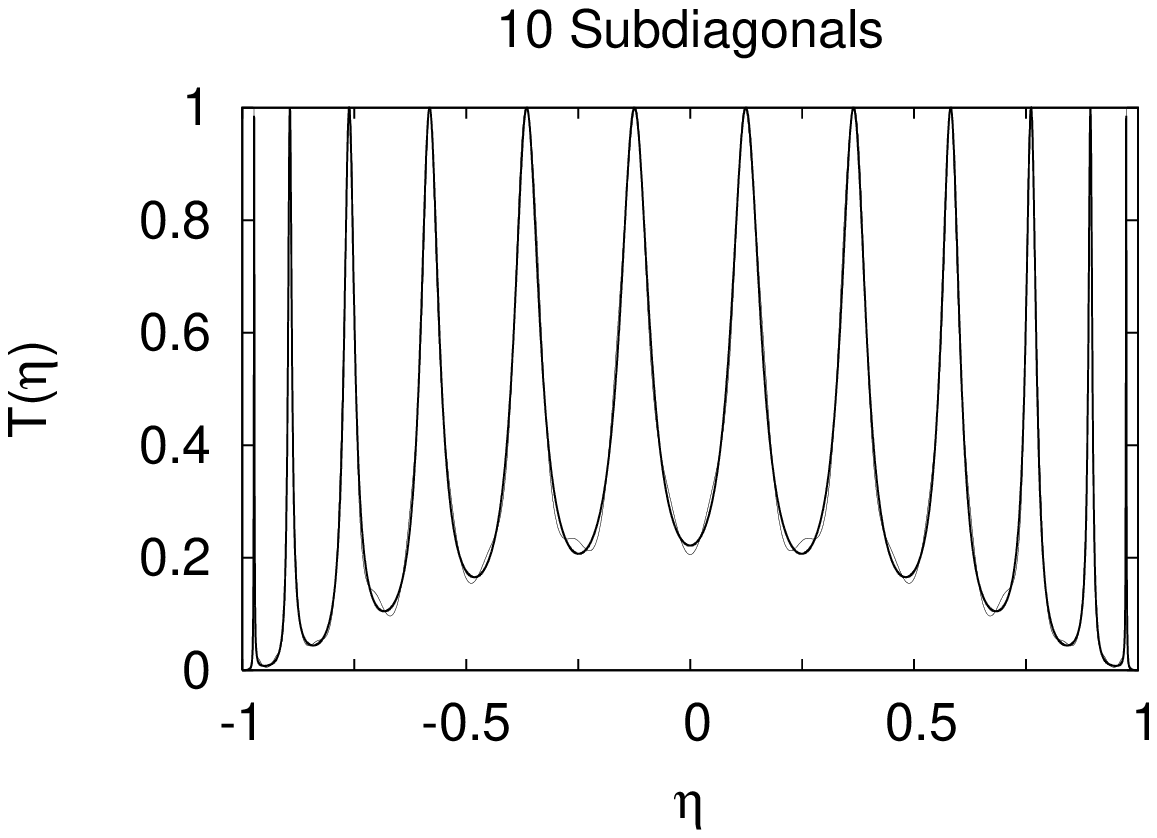}
\hspace{1cm}
\includegraphics[width=7.2cm]{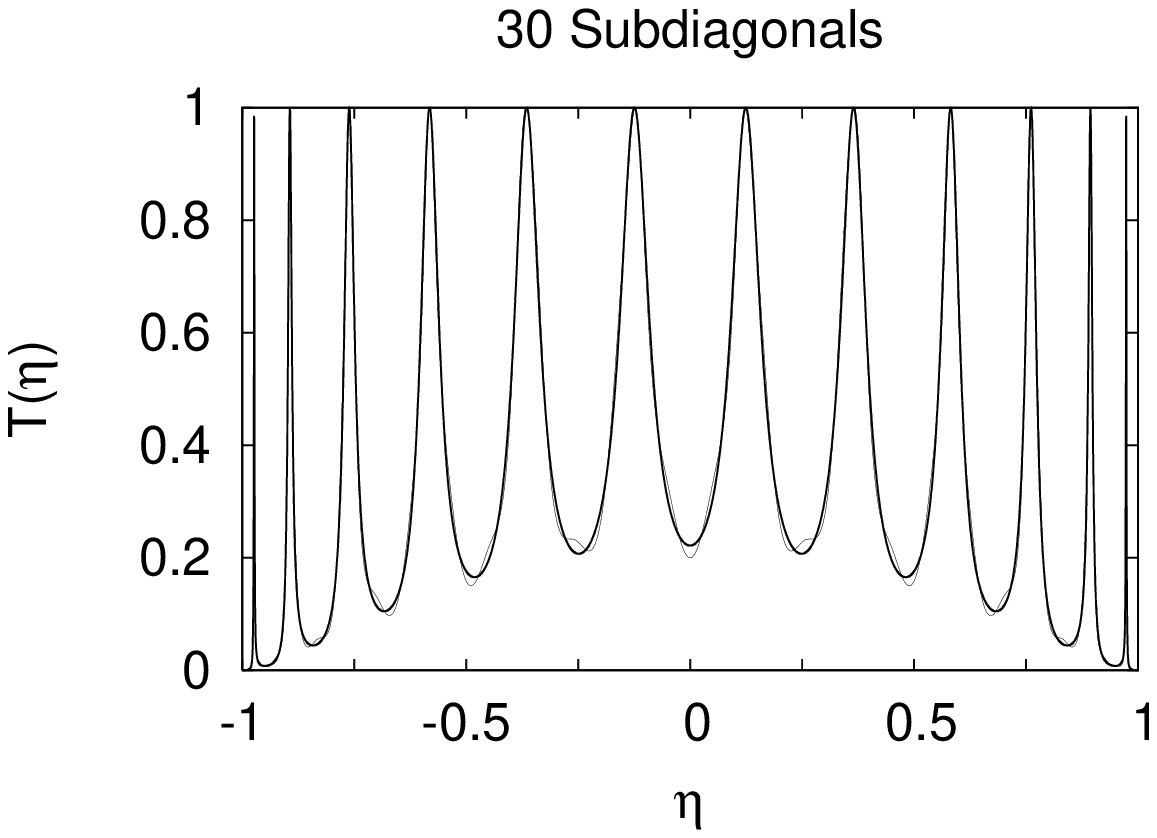}
\caption{Transmission function with $\NL = \NR = 30$ and $N = 12$ for various 
values of $N_S$.  The exact result is shown with the heavy line, while the 
result from our complex potential in a semi-local approximation is given in 
the lighter line. We recall that $\eta = (E - \epsilon_0) / {2 \gamma}$.
\label{fig:LocalityTest}}
\end{figure*}

\section{Discussion and Conclusions}
\label{sec:Conclusions}
We have given a prescription for calculating a complex absorbing potential 
from the self-energy, thus enabling one to describe coupling to continuum 
states in an energy-independent way.  In our test
system, the non-empirical complex potential gives reasonable results 
even at the simplest level of approximation, but by obtaining the potential 
for each electrode individually and using the proper states to construct the 
potential, we can build a non-local complex potential that yields for most 
practical purposes essentially the exact transmission function.

We have several main observations to make from this initial work.  

First of all, it is essential that the complex potential yield not only the 
correct energy levels but also the correct states.  This is not surprising, as 
one function of the complex potential is to force free-particle states to look 
rather like bound states outside of the area of interest.  It is no 
coincidence, then, that inside the area of interest, the states should be 
greatly modified from the original bound states to closely resemble
the resonant eigenfunctions.

Secondly, in cases such as the one we have examined, where there are multiple 
self-energies in the problem, we should construct a different complex 
potential for each self-energy, and combine them to form a total complex 
potential at the end.  While we can get away with a somewhat {\it ad hoc} 
construction if we have some physical insight, the results are nevertheless 
not as good as when we build the complex potential for each self-energy 
separately.

Third, we can, surprisingly, eliminate most of the absorption grid outside the 
subsystem region of primary interest. This was shown with the explicit 
elimination of all electrode atoms in our simple system, and it is apparently 
another consequence of using the correct resonance structure and states. The 
main effect of increasing the number of explicit electrode atoms (which 
presumably decreases the burden on the complex potential) is to improve the 
description at energies off resonance or near the band edge.  The great 
utility of such a significant reduction in the number of required degrees of 
freedom is readily appreciated if one considers that the bulk metal electrodes
can be removed almost completely from electron transport calculations with 
CAPs.

Our complex potential works only imperfectly for the ideal wire.  Again, most 
of the problems are at the band edge, but if we do not include a sufficient 
number of sites, the description even in the middle of the band has some 
unphysical oscillations.  Of course, in such a system, there are no resonances 
and one would not really expect a complex potential to describe the physics 
perfectly anyway.  Nevertheless, convergence of the results can be tested as 
one increases the number of sites explicitly treated.

Finally, our complex potential is strongly non-local, and that non-locality is 
essential in its functioning.  A semi-local approximation is not terrible, and 
works quite well on resonance, but the full non-locality is needed if we are 
to accurately describe energies off resonance.

All of this seems to suggest that, so long as one is careful in how one uses 
a self-energy transformed complex potential of the sort we have introduced, 
one should easily be able to obtain results within the required accuracy
without having to introduce parameters.

A limitation to our approach is the requisite prior knowledge of a
self-energy at some level of theory.  Therefore, our method is ill-suited
to applications for which the self-energy is difficult to construct.
However, there is a wide range of problems that can be considered.
In the simple case that many-body effects are of major concern only within a
subsystem~\cite{JCP06Tom} while the surrounding environment can be adequately
treated at the single-particle level, our self-energy transformed complex
potential should be easy to construct and would allow us to readily describe the
embedded subsystem at a high level of theory.  In particular, for quantum
transport across molecular junctions it has been shown that many-particle
effects~\cite{PRL03DG} can play a critical role,  but including the continuum of
the reservoirs has proven challenging.  Our potentials should enable us to treat
the extended molecule with an accurate wave function approach while still including
a reasonable description of the bulk electrodes.

\section*{Acknowledgment}
We would like to thank Science Foundation Ireland (SFI) for funding this 
work.  One of us (E.H.) was funded through the SFI UREKA program.


\begin{thebibliography}{37}
\bibitem{note1}
If $\ket{\Psi}$ is an eigenvector of a Hermitian operator, then its 
Hermitian adjoint $\ket{\Psi}^{\dagger}$ will be the same as its dual
$\bra{\Psi}$.  Because $H_0 + \Sigma$ is non-Hermitian, the dual of one of its 
eigenvectors will not be the same as its adjoint.  In an abuse of notation, we 
write $\bra{\Psi} = \ket{\Psi}^{\dagger}$ even if $\ket{\Psi}$ comes from a 
non-Hermitian operator.  That is, the bra is always meant to indicate the 
Hermitian adjoint of the ket, rather than its dual.
\bibitem{note2}
That there are no factors of one half in the Green's function is 
because each of $\WL$ and $\WR$ contributes 
$1/2 \; \delta_i - 1/2 \; \mathrm{i} \, \lambda_i$.
\bibitem{note3}
We could of course use $\bar{\W}^{\phi}$ instead, as it is equally good.
\end{thebibliography}
\end{document}